\documentclass[journal]{IEEEtran}

\usepackage{cite}
\usepackage{amsmath,amssymb,amsfonts}
\usepackage{ulem}
\usepackage{algorithmic}
\usepackage{graphicx}
\usepackage{textcomp}
\usepackage{xcolor}

\usepackage{caption,subcaption}
\usepackage{array}
\usepackage{hhline}
\usepackage{multirow}
\usepackage{hyperref}
\usepackage{amsmath, amssymb}
\usepackage{booktabs}

\usepackage{placeins}
\usepackage{tabularx} 
\usepackage{booktabs}
\usepackage{makecell} % 支持 \makecell
\usepackage{arydshln} % 支持虚线 \hdashline
\usepackage{adjustbox}

\def\BibTeX{{\rm B\kern-.05em{\sc i\kern-.025em b}\kern-.08em
    T\kern-.1667em\lower.7ex\hbox{E}\kern-.125emX}}

\begin{document}

\title{Tracing Your Account: A Gradient-Aware Dynamic Window Graph Framework for Ethereum under Privacy-Preserving Services}

% \author{Shuyi~Miao, Wangjie~Qiu, Xiaofan~Tu, Yunze~Li and~Zhiming~Zheng}
% \thanks{Jinchun He, Wangjie Qiu, Qinnan Zhang, and Zhiming Zheng are with the Institute of Artifcial Intelligence, Beijing Advanced Innovation Center forFuture Blockchain and Privacy Computing, Beihang University, Beijing 100191 China, also with Zhongguancun Laboratory, Beijing, China, China (E-mail:hejinchun@buaa.edu.cn; wangjieqiu@buaa.edu.cn; zhangqn@buaa.edu.cn,zzheng@pku.edu.cn).}% <-this % stops a space

\author{Shuyi~Miao, Wangjie~Qiu, Xiaofan~Tu, Yunze~Li, Yongxin~Wen, and Zhiming~Zheng%
\thanks{Shuyi Miao, Wangjie Qiu, Xiaofan Tu, Yunze Li, and Zhiming Zheng are with the Institute of Artificial Intelligence, Beijing Advanced Innovation Center for Future Blockchain and Privacy Computing, Beihang University, Beijing 100191, China (E-mail: shuyimiao@buaa.edu.cn; wangjieqiu@buaa.edu.cn; xiaofantu@buaa.edu.cn; tedlee@buaa.edu.cn; zzheng@pku.edu.cn.) Shuyi Miao, Wangjie Qiu, and Zhiming Zheng are also with Zhongguancun Laboratory, Beijing, China.}
\thanks{Yongxin~Wen is with the No.208 Research Institute of China Ordnance Industries, Beijing, China (E-mail: wyx1142557393@163.com.)}
        
\thanks{Corresponding Authors: Wangjie Qiu.}}
% \author{Shuyi~Miao$^{1,2,3}$ ,
%         Wangjie~Qiu$^{1,2,3*}$,
%         Xiaofan~Tu$^{1,2,3}$,
%         Yunze~Li$^{1,2,3}$,
%         and~Zhiming~Zheng$^{1,2,3}$\\
% $^1$the Institute of Artificial Intelligence, Beihang University, Beijing 100191, China\\ 
% $^2$Beijing Advanced Innovation Center for Future Blockchain and Privacy Computing, \\Beihang University, Beijing 100191, China \\
% $^3$Zhongguancun Laboratory, Beijing, China \\
% \texttt{\{shuyimiao, wangjieqiu, xiaofantu, \}@buaa.edu.cn},\\
% \texttt{{\{zzheng\}@pku.edu.cn},
% % \texttt{dongjin@baec.org.cn}
% }}
% <-this % stops a space

\markboth{Journal of \LaTeX\ Class Files,~Vol.~14, No.~8, August~2021}%
{Shell \MakeLowercase{\textit{et al.}}: A Sample Article Using IEEEtran.cls for IEEE Journals}


\maketitle

\begin{abstract}
% With the rapid evolution of Web 3.0, numerous services have emerged to enhance user privacy and anonymity on public blockchains. However, these services' strong untraceability, particularly mixing services, has also attracted criminals seeking to launder illegal funds. Meanwhile, traditional de-anonymization methods struggle with transactions that lack distinguishing features or have uncertain recipients.

With the rapid advancement of Web 3.0 technologies, public blockchain platforms are witnessing the emergence of novel services designed to enhance user privacy and anonymity. However, the powerful untraceability features inherent in these services inadvertently make them attractive tools for criminals seeking to launder illicit funds. Notably, existing de-anonymization methods face three major challenges when dealing with such transactions: highly homogenized transactional semantics, limited ability to model temporal discontinuities, and insufficient consideration of structural sparsity in account association graphs.
To address these, we propose GradWATCH, designed to track anonymous accounts in Ethereum privacy-preserving services. 
Specifically, we first design a learnable account feature mapping module to extract informative transactional semantics from raw on-chain data. We then incorporate transaction relations into the account association graph to alleviate the adverse effects of structural sparsity. To capture temporal evolution, we further propose an edge-aware sliding-window mechanism that propagates and updates gradients at three granularities. Finally, we identify accounts controlled by the same entity by measuring their embedding distances in the learned representation space.
%实验结果总结句
Experimental results show that even under the conditions of unbalanced labels and sparse transactions, GradWATCH still achieves significant performance gains, with relative improvements ranging from 1.62\% to 15. 22\% in the MRR and from 3. 85\% to 7. 31\% in the $F_1$.

% Next, we model transactions as a heterogeneous dynamic graph and partition them into fixed-length sliding windows. We develop edge-aware graph encoding within each time slice to capture topological relationships. Within each window, we propagate instantaneous gradients between adjacent time slices and ultimately aggregate them to generate adaptive cumulative gradients. Subsequently, the updated window parameters are propagated across windows to maintain temporal continuity and enhance the representation of account nodes.

\end{abstract}

\begin{IEEEkeywords}
Blockchain, Ethereum, Privacy Protection Services, Dynamic Graph Neural Networks, Gradient propagation.
\end{IEEEkeywords}

\bstctlcite{BSTcontrol}
% 双栏单倍行距，字体大小为 10 号
\section{Introduction}
% 区块链-账户地址匿名-吸引犯罪份子-现有去匿名工作
\IEEEPARstart{B}{lockchain} technology \cite{blockchain}, best known for underpinning cryptocurrencies, has been widely adopted across finance and other domains.
However, on public blockchains such as Ethereum, users operate under anonymous alphanumeric addresses. This obscurity complicates regulatory oversight and is often exploited for illicit activities, posing security threats. 
To address these concerns, considerable research has focused on developing account de-anonymization techniques that seek to identify individuals \cite{msy,lindan}, transactions \cite{transaction}, or groups \cite{group} within the blockchain by correlating pseudonymous data with auxiliary information \cite{de-anonymization}. These existing methods typically learn behavioral patterns by modeling observable on-chain transaction paths, thereby enabling relatively effective identification of identity and behavioral inference.

\begin{figure}[!t]
\centering
\includegraphics[width=8cm]{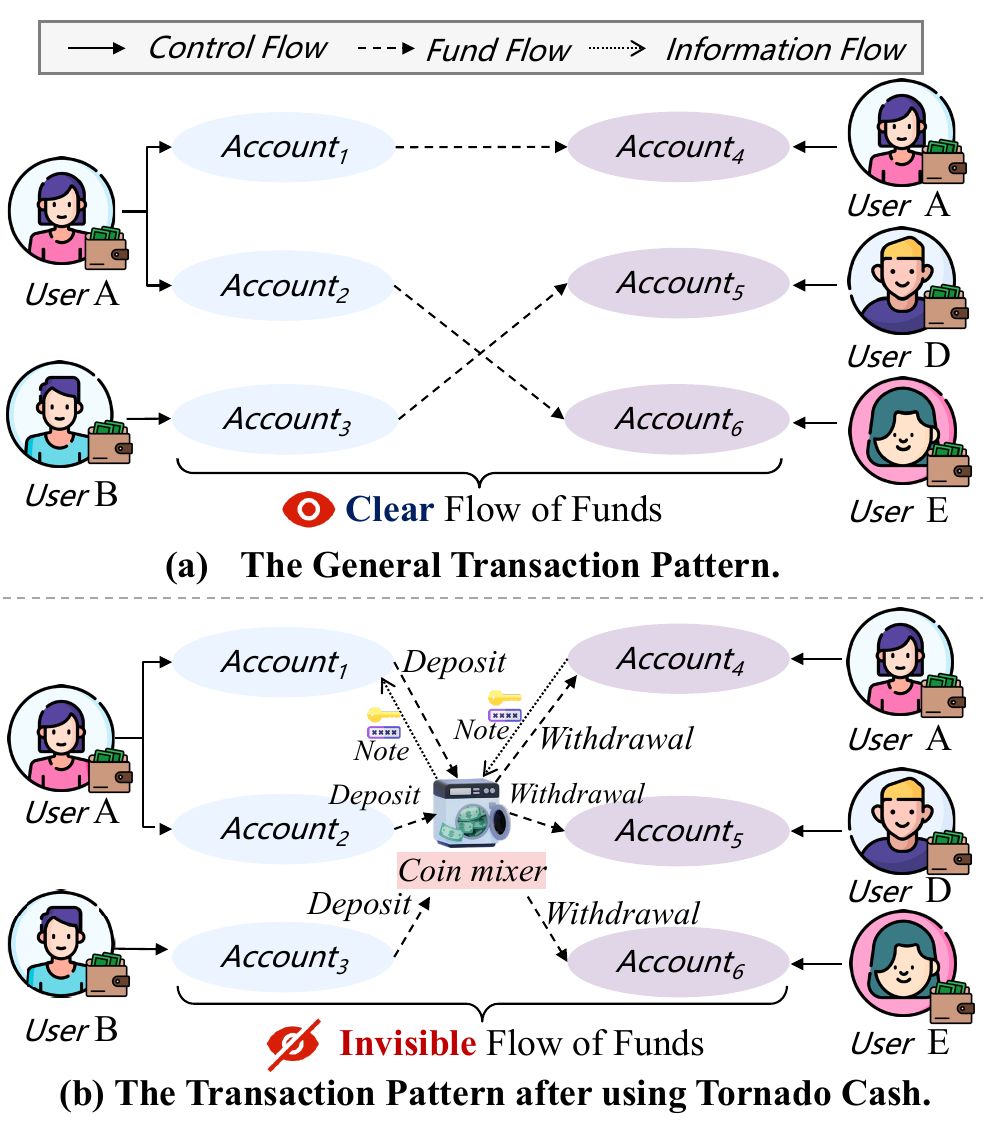}
\caption{Illustration of the Tornado Cash mixing mechanism, contrasting conventional transactions with Tornado Cash transactions to show how deposit and withdrawal addresses are decoupled.}
\label{tcfig}
\end{figure}

% 
%一些隐私保护服务切断了可见的交易路径导致传统方法实效
% However, privacy-preserving technologies in the Web 3.0 ecosystem~\cite{web3}, including mixing services and privacy-enhancing protocols (e.g., Railgun), deliberately obscure or break such paths, thereby fundamentally undermining transaction traceability on public blockchains and rendering traditional approaches ineffective.
% % 引出tornado cash的重要性（为什么要研究），机制
% Among these privacy-preserving technologies, Tornado Cash\footnote{https://tornadoeth.cash} is among the most representative services. As one of the earliest decentralized mixing services deployed on Ethereum, it has long maintained a significant share of usage. A notable case reported by Lianan Tech\footnote{https://trace.lianantech.com/} revealed that, following the WazirX\footnote{https://wazirx.com} breach on July~18,~2024, attackers laundered over \$235~million in stolen assets through Tornado Cash. 
% 
However, privacy-preserving technologies in the Web~3.0 ecosystem~\cite{web3}, including mixing services and privacy-enhancing protocols (e.g., Railgun), deliberately obscure or disrupt such transaction paths.
A representative example is Tornado Cash, one of the earliest decentralized mixing services on Ethereum.
% tonado cash运行机制
As illustrated in Fig.~\ref{tcfig}, Tornado Cash breaks the one-to-one correspondence between deposit and withdrawal transactions. In practice, a \textit{user A} first deposits a fixed-denomination amount into Tornado Cash using \textit{$Account_1$}, which generates a cryptographic commitment associated with a secret deposit note retained off-chain. At a later time, \textit{user A} can authorize a withdrawal to another controlled account \textit{$Account_4$} by presenting a zero-knowledge proof derived from the note.
Importantly, withdrawals are executed from a shared pool and do not reference the original deposit, eliminating deterministic deposit-withdrawal links on-chain. 
This makes the explicit flow of funds impossible to observe, fundamentally undermining transaction traceability on public blockchains.
% and significantly weakening the effectiveness of existing anonymization methods.

\begin{figure}[!t]
\centering
\includegraphics[width=8cm]{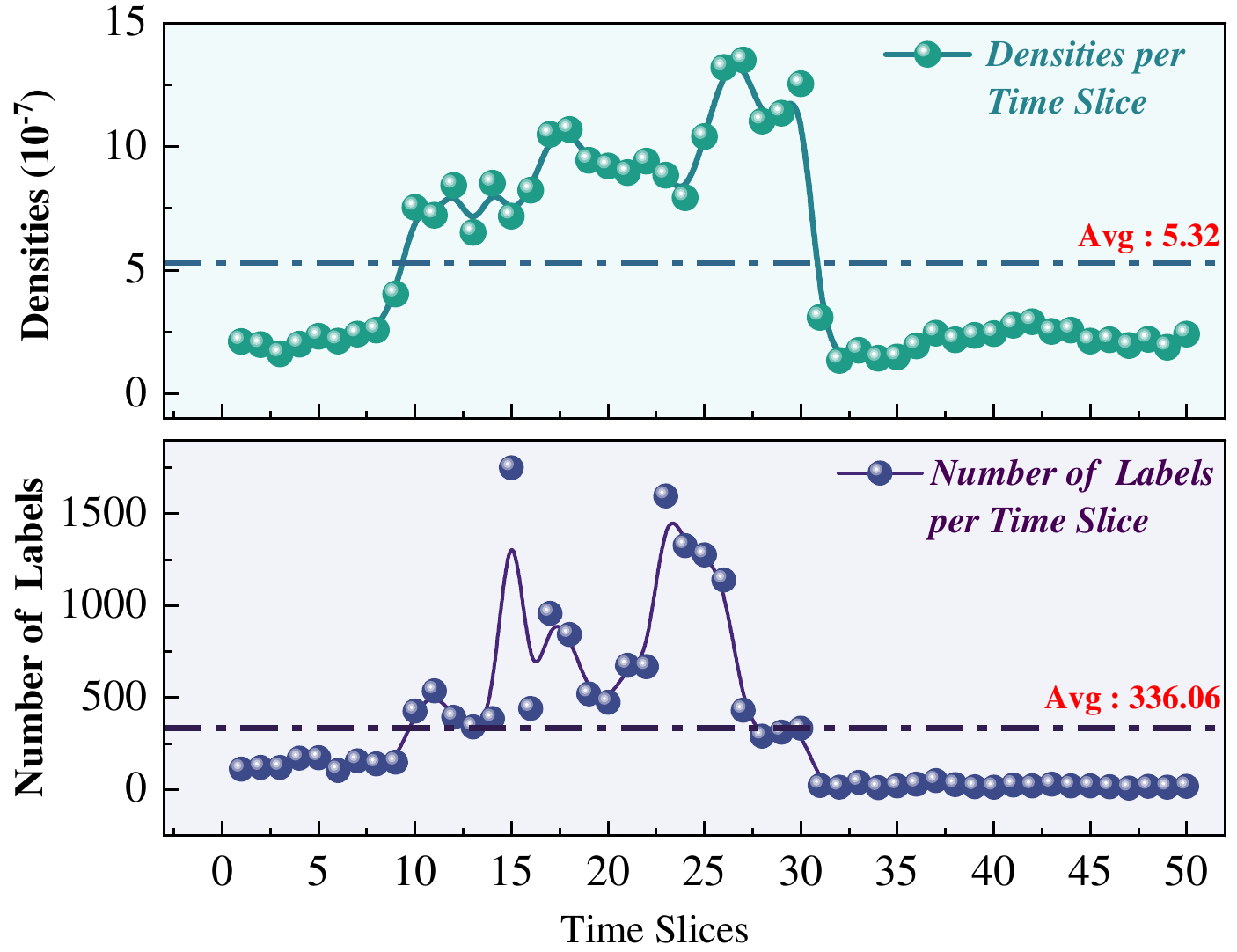}
\caption{Density and number of labels per time slice of the dynamic graph on the transaction-based dataset (50 time slices).}
\label{fig_0}
\end{figure}

To enhance regulatory capabilities for privacy protection in unconventional transaction scenarios, we focus on the \textbf{account tracing} task (i.e., account association detection). This task aims to determine whether multiple accounts appearing in deliberately obfuscated transaction paths are controlled by the same real-world entity, thereby providing an effective means to identify and track illegal money launderers.
%
% However, existing de-anonymization solutions can be broadly categorized into two groups.
However, existing de-anonymization solutions in both categories often struggle to address this task.
\textbf{(i) Heuristic-based methods} rely on handcrafted rules and domain-specific assumptions, which limit their scalability and make them ineffective against adversaries that deliberately evade detection. 
\textbf{(ii) Behavioral feature-based methods} automatically associate accounts by training predictive models to learn latent transaction patterns from blockchain data. While more flexible, behavioral feature-based methods remain insufficient in privacy-preserving environments and exhibit several fundamental challenges (C):
%基于行为特征方法的缺陷
\textbf{(C1) \textit{Homogenized transactional semantics.}} 
In mixing services, transactions are highly homogenized due to fixed denominations and standardized contract interactions. As a result, surface-level behavioral features lack sufficient discriminative power to capture genuine account-control semantics, rendering behavioral similarity a weak and noisy signal for reliable association detection.
\textbf{(C2) \textit{Limited modeling of temporal–structural dynamics.}} 
As illustrated in the lower part of Fig.~\ref{fig_0}, the transaction-based dataset is divided into 50 time slices, across which the account association labels are sparse and unevenly distributed, exhibiting significant variation and abrupt changes in certain periods. However, most existing methods rely on static graph modeling, whereas the few dynamic approaches typically decouple temporal information from graph structure, thereby limiting their ability to capture true temporal dynamics. 
% Account association graphs constructed solely from labelled control relationships are extremely sparse. 
\textbf{(C3) \textit{Structural sparsity of account association graphs.}}
As shown in the upper part of Fig.~\ref{fig_0}, the graph density of the account association graph exceeds the average value of \(5.32 \times 10^{-7}\) only during time slices 10- 30, remaining far below normal graph density standards in most periods. This extreme sparsity severely hinders effective feature propagation and degrades both learning stability and computational efficiency.

Motivated by these challenges, we propose a \uline{\textbf{Grad}}ient-aware dynamic \uline{\textbf{W}}indow graph framework for Ethereum \uline{\textbf{A}}ccount association \uline{\textbf{T}}racking under priva\uline{\textbf{c}}y protection services (\textbf{GradWATCH}). Specifically, for \textbf{C1}, we encode transactions as learnable sequential representations that capture temporal, monetary, spatial, and contextual attributes, and use a reconstruction loss function to ensure semantic alignment between the encoded features and the raw transactions. This design mitigates the tendency of highly homogeneous representation of transaction attributes, providing a more reliable initial representation for account nodes. Second, for \textbf{C3}, by analyzing the mechanisms of Tornado Cash, we exclude other service addresses associated with ambiguous transaction paths, such as Relayer and Proxy. Based on the purified mixing transaction, a dynamic graph of account association driven by mixed transactions (MixTAG) is constructed, enabling learning of transaction-driven account association relationships and mitigating the challenges posed by graph sparsity. Third, for \textbf{C2}, we propose a dynamic heterogeneous graph framework that integrates a sliding window with an edge-aware propagation mechanism.
Finally, account associations are formulated as a link prediction task, in which the likelihood of entity control is computed by measuring the Euclidean distance \cite{euclidean} between node embeddings in the latent space. 
% This end-to-end approach unifies temporal dynamics, topological structures, and behavioral semantics into a cohesive learning paradigm.

In summary, our main \textbf{contributions} are as follows:
\begin{itemize}
\item \textbf{Transaction-driven Heterogeneous Graph Modeling}. We unify account transactions and associations into a graph, eliminating reliance on expert knowledge while boosting graph density.

\item \textbf{End-to-End Account Tracking Paradigm.} By integrating transaction-to-account mapping with an edge-aware window mechanism, our method captures richer transaction semantics and evolutionary evidence to facilitate robust account tracking.

\item \textbf{Empirical Superiority \& Robustness}. 
The experimental results show strong performance across two datasets of different scales, achieving relative improvements of MRR of 1.62\% to 15.22\% and $F_1$ of 3.85\% to 7.31\% compared with baseline approaches. Our implementation is available at: \href{https://github.com/msy0513/GradWATCH}{https://github.com/msy0513/GradWATCH}.

% Evaluations across five metrics (e.g., precision, recall, F1-score) demonstrate state-of-the-art performance. Notably, under extreme label scarcity (10\% training data), GradWATCH exhibits only a 2.3\% performance degradation, underscoring its robustness.
\end{itemize} 

\section{Related Work}
In this section, we first define the address clustering and account tracking tasks. We then provide a detailed review of the research and challenges in detecting tasks in mixing services. Finally, we present the background and current developments in dynamic graph link prediction methods.
\subsection{Address Clustering and Account Tracking}
Based on publicly available on-chain transactions, address clustering \cite{address_cluster} involves analyzing account behavior patterns and classifying addresses into the largest subsets based on whether the same entity controls them. When the focus shifts from the overall control relationship of entities to determining whether two specific account addresses belong to the same entity, the task is called account tracking \cite{watching}. The core of account tracing is the identification of associations between address pairs. It can be viewed as a refinement and supplement to address clustering results, helping to address incomplete or misclassified issues caused by data noise, privacy-preserving technologies, or the complexity of transaction patterns.

\subsection{Account Tracking in Mixing Services}

In Tornado Cash's mixing service, distinct on-chain deposit and withdrawal addresses may correspond to the same entity since withdrawal parties with deposit vouchers can claim funds. Due to the obfuscation of fund flows by mixing services, many illicit activities conceal the sources of funds through these services, leading to the development of various related downstream tasks, such as account tracking \cite{watching,wang2023zkp,bert4eth,shenmeng,mix_aad,StealthLink,HiLoMix}. The solutions to this task are still scarce, but existing solutions can be broadly classified into two categories: \textbf{(1) Heuristic-based methods.} B\'eres et al. \cite{watching} performed address tracing based on temporal activity patterns and transaction fee models. Wang et al. \cite{wang2023zkp} proposed an expanded set of heuristics, leveraging multi-denomination patterns and TORN mining rewards to further uncover address associations within zero-knowledge-based mixing protocols. However, their approach relies on fixed behavioral features of account associations. 
\textbf{(2) Behavioral feature-based methods.}
Hu et al. \cite{bert4eth} introduced a pre-trained Transformer \cite{transformer} for account association detection 
However, its poor performance raises concerns about the method's effectiveness in real-world deployment.
In contrast to using language models, Du et al. \cite{shenmeng} proposed MixBroker, which constructs an account association graph. This approach relies heavily on expert-designed features, and the sparse graph structure, with its limited topology and missing temporal factors, leads to subpar learning and unsatisfactory performance. To mitigate the account association scarcity issue, Che et al. \cite{StealthLink} proposed StealthLink, a cross-task transfer learning framework that leverages invariant features from malicious account detection to enhance mixing traceability. Tu et al. \cite{HiLoMix} proposed HiLoMix, which employs frequency-aware contrastive learning with confidence-based supervision to mitigate graph sparsity and label noise.

\subsection{Dynamic Graph Link Prediction}
% We model mixing transactions as a heterogeneous dynamic graph driven by mixing transactions and perform account association detection tasks based on this model. In this task, accounts are connected through transaction edges generated by transactions, and account association edges are established via entity control relationships. 
Link prediction is a fundamental task in graph and network analysis, which aims to determine whether an edge exists or will form between a pair of nodes in the future \cite{14}. The account tracing task can naturally be formulated as a link prediction problem, where the goal is to predict whether an association edge exists between two account nodes. To address link prediction, researchers have proposed various graph modeling approaches, including graph random walks, static graph neural networks, and dynamic graph neural networks. 

Graph random walk methods, such as DeepWalk \cite{deepwalk} and Node2Vec \cite{node2vec}, learn node embeddings by simulating walks and capturing structural information. Static graph neural networks, including GCN \cite{gcn}, GAT \cite{gat}, GIN \cite{gin}, and GraphSAGE \cite{graphsage}, learn representations on fixed graphs through neighborhood aggregation and have shown effectiveness in link prediction tasks.
Dynamic graphs incorporate temporal information to model structural evolution. In this context, F. Manessi et al. \cite{dgnn} proposed DGNN, which uses snapshot sequences and GRU \cite{gru} units to maintain historical states. A. Pareja et al. \cite{evolveGCN} introduced EvolveGCN-O and EvolveGCN-H, which model temporal changes by evolving GCN weights or node representations. Similarly, P. Goyal et al. \cite{dyngraph2vec} proposed DynGraph2Vec, which generates graph embeddings at each time step and captures temporal dynamics using LSTM \cite{lstm}. Zhu et al. \cite{wingnn} proposed WinGNN, an encoder-free framework that uses randomized sliding windows for efficient dynamic graph learning. These methods are discrete-time dynamic graph models suitable for graphs with clear time intervals. However, these methods are primarily general graph-modeling approaches and exhibit limited learning capabilities when faced with the challenges of structural sparsity and label scarcity commonly encountered in transaction networks under privacy-preserving services. Therefore, there is an urgent need to develop an account tracing approach tailored to dynamic, heterogeneous graphs that incorporates domain-specific transaction characteristics.

\section{Transaction-driven account association graph construction}
\begin{figure*}[!t]
\centering
\includegraphics[width=\linewidth]{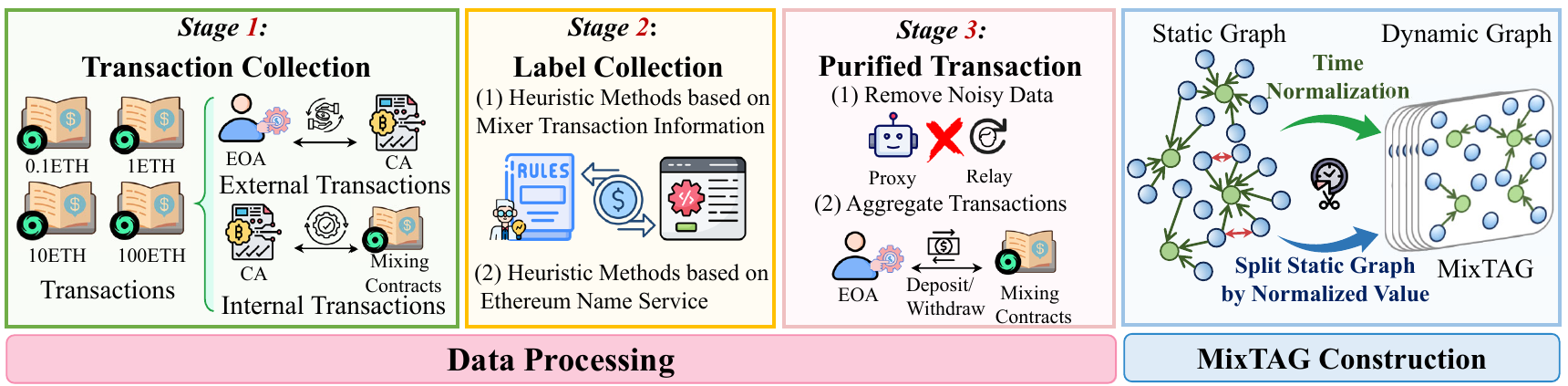}
\caption{Flowchart for constructing MixTAG from mixing transactions and account association labels.}
\label{fig_1}
\end{figure*}

In this section, we will detail the process of generating MixTAG by mixing transaction and account association labels in two parts: data processing and graph construction. The corresponding flow is shown in Fig. \ref{fig_1}.

\subsection{Data processing}
\label{sec:section3.1}

\textbf{Stage 1: Ethereum Mixing Transaction Data Collection.} We collect raw transaction data from the Ethereum blockchain explorer Etherscan\footnote{https://etherscan.io/} related to mixing service contracts with denominations of 0.1 ETH, 1 ETH, 10 ETH, and 100 ETH. Since mixing transactions may involve other services and contracts, such as Relayer or Proxy, a mixing transaction can contain multiple internal transactions. In this process, we retain all external and internal transactions associated with mixing services. External transactions are on-chain operations directly initiated by Externally Owned Accounts (EOAs), while internal transactions are automatically executed message calls triggered between smart Contract Accounts (CAs) or within a contract. Notably, internal transactions do not have their own transaction hashes; instead, they inherit the hash of the external transaction that triggered them.

\textbf{Stage 2: Ethereum Account Association Label Collection.} Inspired by heuristic label generation methods from prior research \cite{watching,bert4eth,shenmeng}, we update and consolidate an existing labeled dataset. The label sources fall into two categories: (1) Heuristic methods based on mixing transaction information: Associations are inferred by analyzing specific patterns in the last nine digits of gasprice values and the chronological sequence of deposit/withdrawal transactions (e.g., time intervals between deposits and withdrawals). (2) Heuristic methods based on Ethereum Name Service (ENS): Associations are identified by tracking ENS domain transfers (via the setOwner function) and subdomain creations (via the setSubnodeOwner function). For instance, if \textit{$account_1$} transfers an ENS domain ownership to \textit{$account_2$} or creates a subdomain for account \textit{$account_2$}, \textit{$account_1$} and \textit{$account_2$} are labeled as linked accounts controlled by the same entity.

\textbf{Stage 3: Purified Mixing Transactions.} Tornado Cash enhances transaction anonymity by offering auxiliary services such as Proxy and Relayer, which allow accounts to deposit and withdraw funds without directly interacting with the mixing contract on-chain. For instance, when an account deposits via Proxy, an internal transaction is automatically triggered, preventing direct interaction between the account and the contract. Similarly, during withdrawals, accounts can use both Proxy and Relayer, where the Relayer covers the withdrawal amount upfront and charges a fee, generating an additional internal transaction. While these services enhance privacy, they also make it significantly more challenging to trace illicit transactions, such as those related to money laundering. Incorporating these extraneous transactions into account association analysis introduces significant noise, complicating representation learning and inference. To address this, we filter out irrelevant internal transactions, retaining only those that directly reflect account to mixing contract interactions, thereby creating a “purified” dataset of internal and external transactions.

\begin{figure}[!t]
\centering
\includegraphics[width=\linewidth]{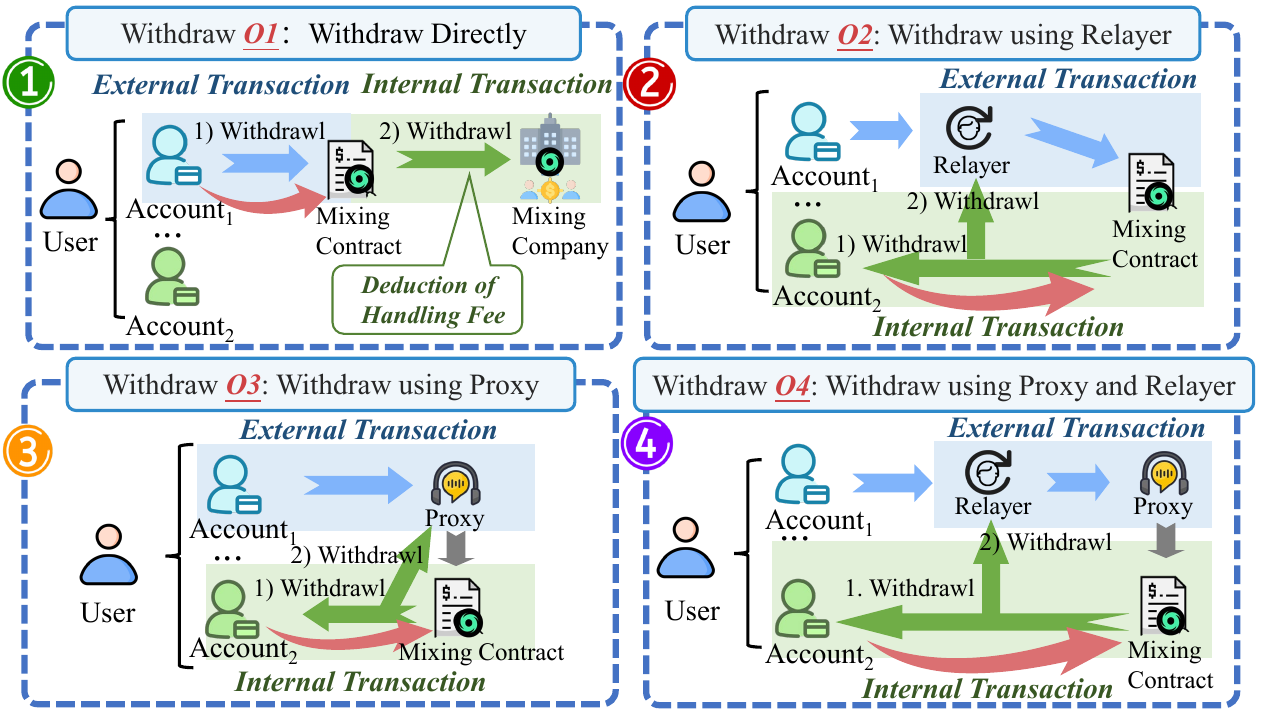}
\caption{The diagram illustrates all possible withdrawal processes. In the figure, \textit{1) withdrawal} represents the transaction where funds are withdrawn to the account, while \textit{2) withdrawal} indicates the transaction used to pay the withdrawal fee. The pink arrows indicate the sender and receiver in the purified transaction.}
\label{fig_2}
\end{figure}

\begin{figure}[!t]
\centering
\includegraphics[width=\linewidth]{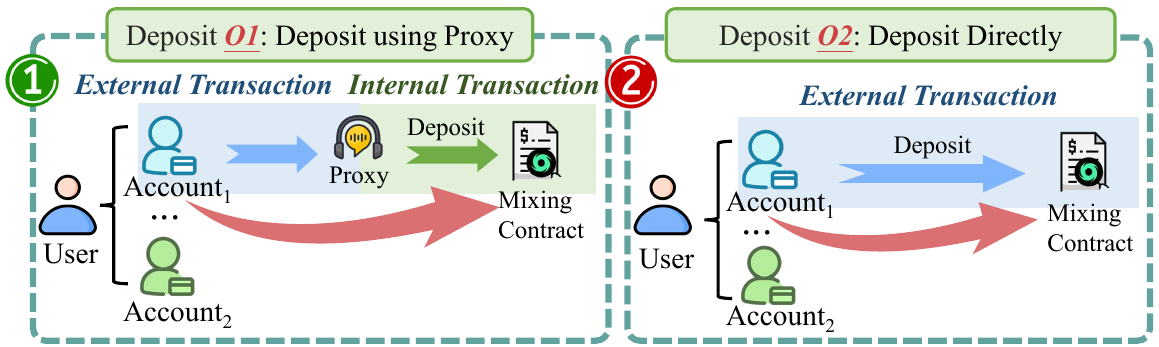}
\caption{All possible deposit process flows using mixing services. The pink arrows indicate the sender and receiver in the purified transaction.}
\label{fig_3}
\end{figure}

We systematically analyze all possible deposit and withdrawal workflows in mixing services to achieve this purification, as illustrated in Fig. \ref{fig_2} and \ref{fig_3}. It is important to note that our dataset includes only transactions directly involving the mixing contract. In certain cases, such as (3) and (4) in Fig. \ref{fig_2}, the external transaction is not directly associated with the mixing contract, resulting in only internal transaction records being available. When processing withdrawal transactions, we first classify them based on whether the transaction hash is linked to an external transaction. \textbf{For withdrawals with external transactions}, we classify a transaction as a withdrawal if it has exactly two internal transactions, e.g., (1) and (2) in Fig. \ref{fig_2}. The actual withdrawal \textit{$account_1$} and its interaction with the mixing contract are then extracted and retained in the purified dataset, where they are labeled as withdrawal transactions highlighted by pink arrows in Fig. \ref{fig_2}. \textbf{For withdrawals without external transactions}, we identify the \textit{$account_2$} as the withdrawal account and the receiver as the mixing contract, ensuring these transactions are also retained and labeled accordingly (3) and (4) in Fig. \ref{fig_2}. This purification process eliminates irrelevant interactions, improving the accuracy of subsequent de-anonymization analysis.

\textbf{For deposit transactions}, only the two scenarios shown in Fig. \ref{fig_3} exist, with the primary difference being whether the Proxy service is used. For deposits made via Proxy, as shown in Fig. \ref{fig_3} (1), we extract the sender of the external transaction (\textit{$account_1$}) and the mixing contract information from the internal transaction, record them in the purified transaction data, and label them as deposit transactions indicated by the pink arrows in the Fig. \ref{fig_3}. In contrast, for deposits made directly from the \textit{$account_1$} as shown in Fig. \ref{fig_3} (2), the corresponding external transaction is directly recorded in the purified transaction data and similarly labeled as a deposit transaction. Through the above steps, the obtained purified transactions exclude Proxy and Relayer services. Additionally, by analyzing and merging transactions, the new dataset explicitly includes the initiator of each transaction and the corresponding mixing contract, eliminating noise and interference introduced by other services.

\subsection{MixTAG Construction}
In this section, we aim to construct a graph using the purified transaction obtained in the section~\ref{sec:section3.1}. Formally, we model the purified transaction as a heterogeneous graph $\mathcal{G}=\{\mathcal{V},\mathcal{E},\mathcal{T_V}, \mathcal{T_E},\mathcal{X}, \mathcal{R}, Y\}$ driven by mixing transactions, where $\mathcal{V}=\{v_{i}\}_{i=1}^{n}$ represents the set of account nodes and $\mathcal{E}=\{e_{i}\}_{i=1}^{m}$ represents the set of edges. $\mathcal{T_V}=\{v_{_e},v_{_c}\}$ represents the set of node types where $v_{_e}$ and $v_{_c}$ denote the EOAs and the CAs, respectively. $\mathcal{T_E}=\{e_{_a},e_{_t}\}$ represents the set of edge types where $e_{_a}$ and $e_{_t}$ denote the account association edge and the account transaction edge, respectively. Moreover, the type of node $v_i$ and each edge $e_{i}$ satisfies $t_{v_{i}}\in \mathcal{T_{_{V}}}$ and $t_{e_{i}}\in \mathcal{T_{_{E}}}$. $\mathcal{X}\in\mathbb{R}^{n\times d_1}$ is the node feature matrix where $n$ is the number of nodes and $d_1$ is the feature matrix dimension. $\mathcal{R}\in\mathbb{R}^{m\times d_2}$ is the edge feature matrix where $m$ is the number of edges and $d_2$ is the dimension of the edge feature matrix. Our task is to determine whether there is an association between two nodes in MixTAG. Formally, for nodes $v_i$ and $v_j$, if they are controlled by the same real-world entity, we define an edge $e_k=(v_i, v_j)$ and assign the label $y_{i,j}=1$. Then, we normalize the transaction time $T$ in the purified transaction. Specifically, assuming that the execution time of the $k$-th mixing transaction is $t_k$, its normalized evolutionary time can be defined as follows:
\begin{equation}
  \ T_{k}=\frac{t_{k}-t_{min}}{t_{max}-t_{min}},
\end{equation}
where $t_{min}$ and $t_{max}$ represent the minimum and maximum transaction times, respectively. Based on the evolutionary time of transactions, we partition the static graph $\mathcal{G}$ into $t$ time slices. The $k$-th time slice can be denoted as $G_k=\{\mathcal{V},\mathcal{E}_k,\mathcal{T_V}, \mathcal{T_E},\mathcal{X}, \mathcal{R}_k, T_k, Y_k\}$, where $\mathcal{E}_k$ includes not only the mixing transaction edges occurring in the 
$k$-th time slice and the account association edges inferred from the account withdrawal times.
% For an association edge between $i$-th account and $j$-th account, we assign it to the time slice corresponding to the withdrawal transaction of the two accounts. This ensures that the model accurately learns the relationship between deposits and withdrawals, making it more aligned with the characteristics of the mixing scenario.

\begin{table}[h]
    \centering
    \fontsize{8.5}{11}\selectfont 
    \caption{The main notation used in this paper.}
    \begin{tabularx}{\columnwidth}{lX} % X 让第二列自动扩展
        \toprule
        \midrule
        Symbol & Description \\
        \midrule
        $\mathcal{G}$ & MixGraph. \\
        $\mathcal{G}_k$ & MixGraph for the $k$-th time slice. \\
        $v, \mathcal{V}$ & Node (account), node set \\
        $e, \mathcal{E}$ & Edge (transaction or account association), edge set. \\
        $t_{v}, \mathcal{T_V}$ & Node types, node types set. \\
        $t_{e}, \mathcal{T_E}$ & Edge types, edge types set. \\
        $x, \mathcal{X}$ & Node feature, node feature matrix. \\
        $r, \mathcal{R}$ & Edge feature, edge feature matrix. \\
        $t, T$ & Transaction time, transaction time set. \\
        $y, Y$ & Account association label, label set. \\
        \hline 
        $s_{i,j}, S_{i}$ & The $j$-th transaction sequence and set for the $i$-th account. \\
       $a_{i,j}$ & Numerical features in the $j$-th transaction of the $i$-th account.\\
       $c_{i,j}$ & Categorical features in the $j$-th transaction of the $i$-th account. \\
       $t_{i,j}$ & Temporal features in the $j$-th transaction of the $i$-th account. \\
       $p_{i,j}$ & Position features in the $j$-th transaction of the $i$-th account. \\
       $H_{i,j}$ & Representation of the $j$-th transaction for the $i$-th account. \\
       $\hat{M}_j$ & The message received by the $j$-th node. \\
        $z_{i}$ & The embedding representation of the $i$-th node. \\
        $Z$ & The set of node embeddings.\\
        $\nabla \mathcal{L}_{G_t}^{w_i}$  & The instantaneous time slice gradient of the $i$-th window.\\
        $\nabla \hat{\mathcal{L}}_{G_{t+1}}^{w_i}$  & The cumulative time slice gradient of the $i$-th window.\\
        $\nabla\hat{\mathcal{L}}^{w_i}$  & The $i$-th window gradient.\\
        \bottomrule
     \midrule
    \end{tabularx}
    \label{tab:notations}
\end{table}

\section{Method}
\subsection{Overview}
\begin{figure*}[!t]
\centering
\includegraphics[width=\linewidth]{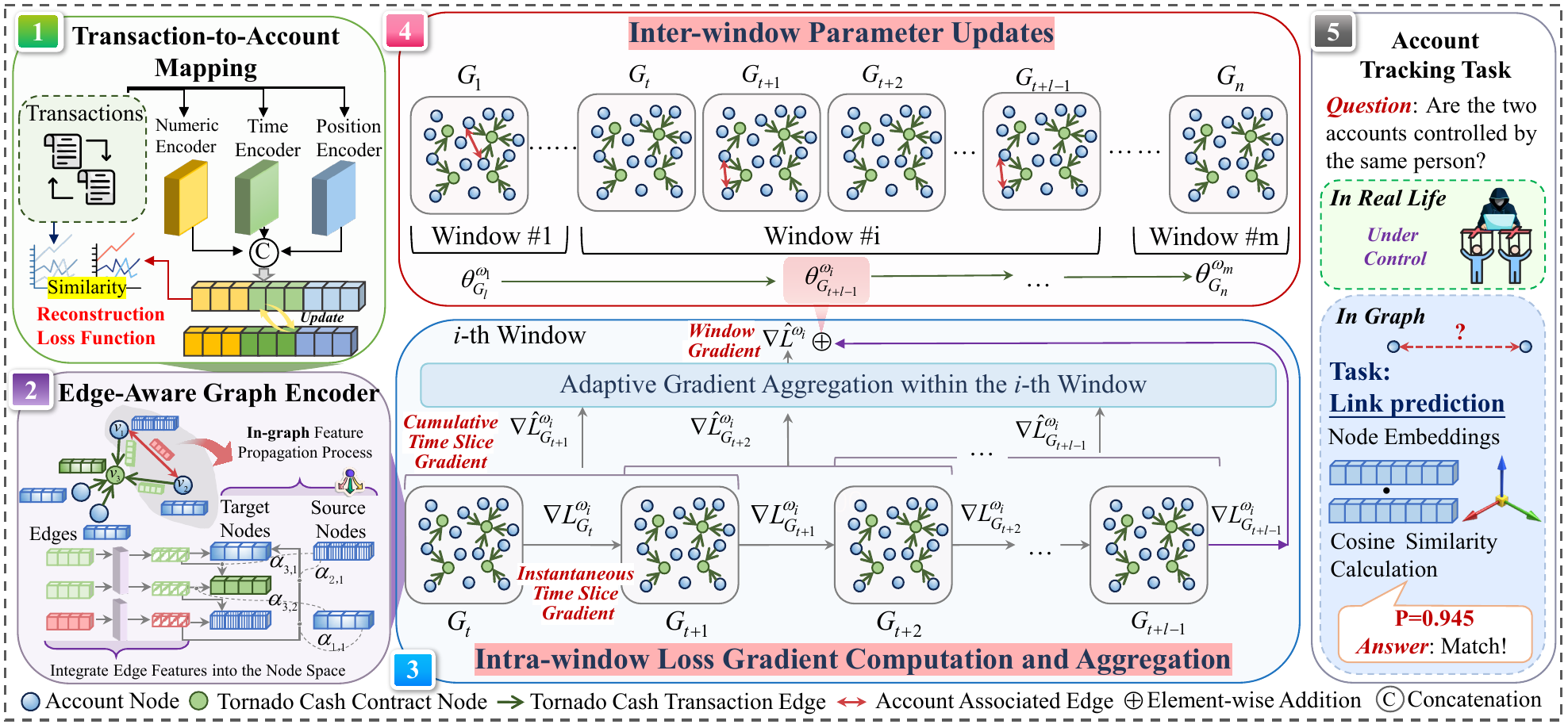}
\caption{The overview of our GradWATCH.}
\label{fig_4}
\end{figure*}

As shown in Fig. \ref{fig_4}, GradWATCH consists of five modules. 
(1) Transaction-to-Account Mapping: This module processes raw transactions to generate initial node embeddings with numerical, temporal, and positional features. A reconstruction loss ensures that embeddings accurately reflect transaction data.
(2) Edge-Aware Graph Encoding: The remaining raw transaction details (excluding those used in (1)) are treated as edge features, and an edge-aware graph convolutional network is applied to update node representations.
(3) Intra-Window Gradient Computation: Multiple consecutive dynamic time slices form a sliding window. Within each window, we compute the instantaneous gradient for the current slice, the cumulative gradient over the window, and the window gradient.
(4) Inter-Window Parameter Update: Model parameters are updated using the window gradient and the instantaneous gradient from the last slice of the current window. This dual-layer mechanism captures both long-term dependencies and short-term variations for robust updates.
(5) Account Association Prediction: We model account association detection as a graph link prediction task, computing the probability of an edge existing between account nodes using the updated embeddings.
By integrating these five modules, GradWATCH effectively captures dynamic account relationships in mixing transactions, enabling high-accuracy link prediction.

\subsection{Transaction-to-Account Mapping Module}
\textbf{\textit{Step 1: Transaction Representation Generation.}} Transaction data can be categorized into four types based on data characteristics: numerical, categorical, temporal, and spatial. For each account, every transaction is classified into its respective data type. For example, the basic features of the $i$-th account's $j$-th transaction sequence $s_{i,j}=\{a_{i,j}, c_{i,j},...\}$, such as numerical and categorical features $a_{i,j}$ and $c_{i,j}$, can be encoded using a numerical encoder as follows:
\begin{equation}
    NE(s_{i,j}) = \text{Concat}(W_a \cdot a_{i,j} + b_1, W_c \cdot c_{i,j} + b_2) + \varepsilon,
\end{equation}
where ${W}_a \in \mathbb{R}^{d_1 \times p}$ and ${W}_c \in \mathbb{R}^{d_2 \times q}$ are the mapping matrices of categorical and numerical features, respectively, $b_1$ and $b_2$ are bias terms. To alleviate the problem of hard boundaries between categories in the representation and to simulate the variability of transaction numerical data in reality, we introduce Gaussian noise $\varepsilon \sim \mathcal{N}(0, \sigma^2)$ to add noise, improving the robustness and generalization ability of the numerical encoder. $d = d_1 + d_2$ denotenabling the model to better perceive transaction timeme information of transactions $t_{i,j}$, we use a relative time encoder based on sine and cosine functions \cite{2,1,3}, which maps time values to a high-dimensional embedding space, allowing the model to better perceive transaction time information. It is expressed as follows:

\begin{equation}
TE(s_{i,j}) =
\begin{cases}
\cos\left(\frac{t_{i,j}}{10000^{(z-1)/d}}\right), & \text{if } z \text{ is odd} \\
\sin\left(\frac{t_{i,j}}{10000^{z/d}}\right), & \text{otherwise}
\end{cases}
\end{equation}
where $z$ represents the dimension index, which is used to incrementally generate the encoding for each dimension. Since different accounts may engage in multiple transactions with the mixing service, account representation should incorporate positional encoding based on the relative positions of transactions within each account's transaction history. The specific implementation is as follows:
\begin{equation}
    PE(s_{i,j})=W_p \cdot p_{i,j},
\end{equation}
where $p$ is the position index and ${W}_p \in \mathbb{R}^{d \times T}$ denotes the position encoding matrix. Therefore, by concatenating the encoding results of each transaction's input value, time, and positional encoders, the comprehensive representation of the $j$-th transaction can be obtained:
\begin{equation}
    H_{i,j} = NE(s_{i,j}) \| TE(s_{i,j}) \| PE(s_{i,j}).
\end{equation}

\textbf{\textit{Step 2: Similarity Constraint.}} After generating the transaction representation, we introduce an autoencoder-based reconstruction loss function to optimize the embeddings of the transactions. Specifically, our reconstruction loss comprises four components, each corresponding to a different encoder. First, we use cross-entropy loss to evaluate the reconstruction of transaction categories, as it effectively measures the accuracy of binary transaction type predictions. Second, for the transaction value encoder, we use Mean Squared Error (MSE) as the loss function to ensure the learned embeddings accurately reflect the distribution of numerical data. Similarly, MSE loss is used to preserve spatial and temporal information for transaction location and time encoders. 
% The final reconstruction loss is the combination of these four components:

\textbf{\textit{Step 3: Account Aggregation Representation.}} 
Furthermore, we aggregate the embeddings of all transactions related to the $i$-th account $v_i$, and the initial representation for the $i$-th account is given by:

\begin{equation}
    x_i = \frac{1}{|T_i|} \sum_{j \in T_i} H_{i,j},
\end{equation}
where $T_i$ is the set of all transactions associated with the $i$-th account $v_i$ and $\left| T_i \right|$ is the number of transactions in the set.

\subsection{Edge-Aware Graph Encoding}
In the MixTAG, each node \( v_i \) have features \( x_i \), which are connected to the feature \( x_j \) of its neighboring node \( v_j \) through an edge \( e_{i,j} \). Due to the dimensional differences between the nodes and edges, effective interaction and fusion in different spaces are challenging. Therefore, we use a linear layer to map the edge \( e_{i,j} \) to the same dimension as the node features, obtaining a new edge representation of the same dimension:
\begin{equation}
    \tilde{e}_{i,j} =  W_e e_{i,j} + b_{e},  
\end{equation}
where \( W_e \) is the weight and  and \( b_e \) is bias terms. Previous studies have typically propagated and updated node information equally through edges in graphs. However, in MixTAG, in addition to transaction edges, there are also account association edges. Therefore, it is necessary to design distinct message-propagation mechanisms for different edge types.

To address this, we improve the GCN's propagation function by integrating edge features and an attention mechanism. This enhancement enables MixTAG to capture critical transaction context information and account association information more effectively in MixTAG, which involves multiple inputs and outputs.
Specifically, we use a nonlinear activation function and a linear transformation to compute graph messages, allowing messages to dynamically allocate weights while comprehensively considering the source node $v_i$, target node $v_j$, and edge information $e_{i,j}$:
\[
m_{i,j} = Sigmoid \left( W_m e_{i,j} + b_m \right) \cdot x_i + \tilde{e}_{i,j} + x_j,
\]
where \( W_m \) and \( b_m \) are the weight and bias terms, respectively. Next, we aggregate all the messages received by the target node  $v_j$:
\begin{equation}
    M_j = \sum_{i \in \mathcal{N}(j)} m_{i,j},
\end{equation}
where ${\mathcal{N}(j)}$ be the set of neighbors of the target node $v_j$. Considering that the MixTAG contains two types of edges: account association edges and transaction edges, the message aggregation across edge types is further formulated as:
\begin{equation}
    \hat{M}_j = \sum_{k \in T_e} M_j^k,
\end{equation}
where $T_e$ be the set of all edge types. Finally, we update the node representation of the target node using the aggregated message:
\begin{equation}
    \hat{x_j} = W_n \cdot \left(\frac{\hat{M_j}}{\|\hat{M_j}\|_2 + \epsilon}\right) + b_n,
\end{equation}
where $W_n$ and \( b_n \) are the weight and bias terms. $\|\cdot\|_2$ represents the L2 norm and $\epsilon$ is a constant to ensure numerical stability.

\subsection{Intra-Window Loss Gradient Computation and Aggregation}
% Traditional static GNNs perform supervised learning on a single graph, calculating the model's loss based on the training data. This loss is then used to update the GNN layers and other learnable parameters via backpropagation. Such an approach learns solely within a single graph and cannot capture the dynamic information across multiple time steps, making it unsuitable for dynamic graphs. Furthermore, some early dynamic graph learning methods add temporal encoders (e.g., GRU, Transformer) on top of the static features captured by static GNNs to encode dynamic features across time steps. However, this strategy introduces more model parameters, increasing training costs and complexity, and is prone to overfitting. In this work, we consider a scenario in which mixing contracts engage in transactions with a large number of users over time. Thus, we aim to use a dynamic GNN that requires fewer model parameters and converges quickly without relying on temporal encoders.

Following the randomized sliding-window mechanism proposed in \cite{wingnn}, we define a window as a collection that spans $l$ consecutive time slices. The window slides forward with a step size $s$, a stochastic variable, meaning that each step size is a random value smaller than the window length $l$. Consequently, the starting position of the next window must always be within the previous window, while its ending position extends beyond the previous window. 
Next, we detail the parameter update mechanism within the window. Inspired by meta-learning strategies \cite{metalearning, metalearning2}, our approach aims to learn from multiple tasks to rapidly adapt to new ones, making it particularly suitable for scenarios with sparse labels. Within each sliding window, we not only compute the loss gradient for the current time slice (that is, \textbf{instantaneous time slice gradient}) but also accumulate and compute the gradient between adjacent time slices (that is, \textbf{cumulative time slice gradient}). For the entire window, we also aggregate multiple cumulative time slice gradients to generate the window gradient, that is, \textbf{window gradient}).
% The aggregated gradient of the window is used to update the model parameters, resembling the inner loop of meta-learning, where each time slice is treated as a training task, and the learned knowledge is propagated to the next time slice.

% % 4.28改到这里
% Subsequently, we update the model parameters using the cumulative time slice gradients from the previous window along with the instantaneous time slice gradient of the last time step in the current window. This process is analogous to the outer loop in meta-learning, where the model parameters are progressively refined by jointly optimizing across all tasks (i.e., all time slices within the window).

% \subsubsection{Calculation of instantaneous time slice gradient}
\textbf{\textit{(1) Calculation of instantaneous time slice gradient.}}
The dynamic graph $ G_t $ at time slice $ t $, after being encoded through $ L $ layers of edge-aware graph convolution, $i$-th node can be represented as:
\begin{equation}
h_{i,t}^{L} = \hat{A} h^{L-1}_{i,t} W_r.
\end{equation}
Next, within this time slice, an MLP is used to predict the probability of an account association edge between node $v_i$ and node $v_j$:
\begin{equation}
\hat{y}_{i,j}^t = \text{MLP}(h_{i,t}^L \, || \, h_{j,t}^L),
\end{equation}
where the symbol $ || $ represents the concatenation operation. Therefore, within a time slice, based on the label data $ y_{i,j}^t \in Y^t $, we can compute the training loss for the existence of an account association edge between nodes $ v_i $ and $ v_j $ using the cross-entropy loss as follows:
\begin{equation}
\mathcal{L}_{G_t}^{w_i} = - \sum y_{i,j}^t \log \hat{y}_{i,j}^t + (1 - y_{i,j}^t) \log (1 - \hat{y}_{i,j}^t).
\end{equation}
We define the $i$-th window between time slices $G_t$ and $G_{t+l}$ as $w_i$ and the symbol $\mathcal{L}_{G_t}^{w_i}$ represents the loss for time slice $G_t$ in the $i$-th sliding window. Furthermore, we compute the instantaneous time slice gradient of the edge-aware graph convolution parameters $\theta_t^{w_i}$ for the corresponding time slice as follows:
\begin{equation}
\nabla \mathcal{L}_{G_t}^{w_i} \simeq \frac{\partial \mathcal{L}_{G_t}^{w_i}}{\partial \theta_t^{w_i}}.
\end{equation}
Subsequently, we use this gradient to update the model parameters for the next time slice $G_{t+1}$:
\begin{equation}
\theta_{t+1}^{w_i}\longleftarrow(\nabla\mathcal{L}_{G_t}^{w_i})^\tau+\theta_t^{w_i},
\end{equation}
where $\theta_t^{w_i}$ is the universal learning rate. 

\textbf{\textit{(2) Calculation of Cumulative time slice gradient.}}
To capture periodic dependencies among multiple time slices within the same window, we introduce a cumulative time slice gradient calculation mechanism, which uses parameters from the current time slice $G_t$ to predict the results of the subsequent time slice $G_{t+1}$. The cumulative time slice gradient on $G_{t+1}$ is calculated as follows:

\begin{equation}
\nabla \hat{\mathcal{L}}_{G_{t+1}}^{w_i} \simeq \frac{\partial \mathcal{L}_{G_{t+1}}^{w_i}}{\partial \theta_t^{w_i}}.
\end{equation}
Through this iterative process, once the gradient calculations are completed from the first to the last slice within the $i$-th sliding window, the window accumulates a total of $l-1$ loss gradients.

\begin{figure}[!t]
\centering
\includegraphics[width=\linewidth]{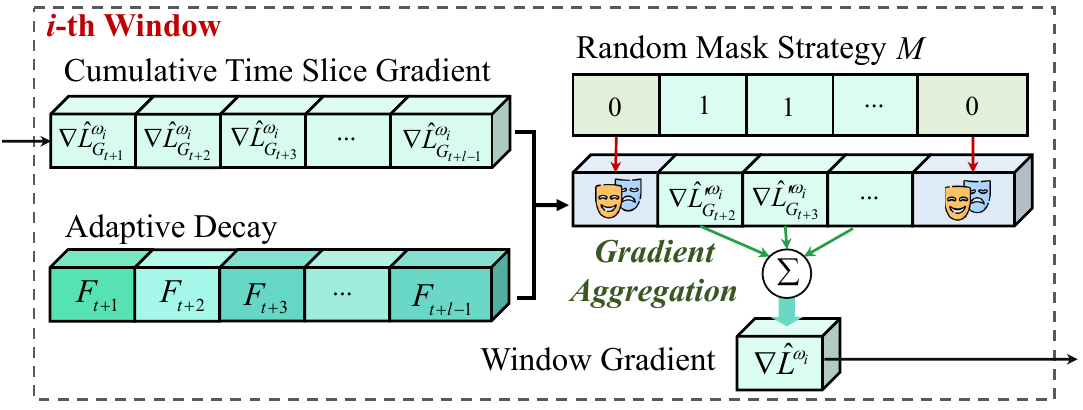}
\caption{The calculation process of the window gradient is adaptively calculated from the cumulative time slice gradients.}
\label{fig_5}
\end{figure}

% \subsubsection{Calculation of Cumulative time slice gradient}
\textbf{\textit{(3) Calculation of window gradient.}}
These $l$ loss gradients need to be aggregated adaptively to facilitate the comprehensive adjustment of the model parameters, as illustrated in Fig. \ref{fig_5}. This can be formally expressed as:

\begin{equation}
\nabla \hat{\mathcal{L}}^{w_i} = \phi \left( \nabla \hat{\mathcal{L}}_{G_{t+1}}^{w_i}, \ldots, \nabla \hat{\mathcal{L}}_{G_{t+l-1}}^{w_i} \right),
\end{equation}
where $\nabla \hat{\mathcal{L}}^{w_i}$ represents the $i$-th window gradient, and $\phi$ denotes the adaptive aggregation function. Unlike the conventional summation approach, to enhance the robustness of dynamic learning, we assign an adaptive decay factor to the gradient values of each time slice within the window. For example, the cumulative time slice gradient $\nabla\hat{\mathcal{L}}_{G_{t+1}}^{w_{i}}$ at time slice $G_{t+1}$ can be assigned a decay factor $F_{t+1}$, and the corresponding calculation can be expressed as:
\begin{equation}
\nabla\hat{\mathcal{L}^{\prime}}_{G_{t+1}}^{w_{i}}=F_{t+1}\odot\nabla\hat{\mathcal{L}}_{G_{t+1}}^{w_{i}},
\end{equation}

\begin{equation}
F_{t+1}=-\frac{\tau}{\sqrt{\delta+r_{t+1}}},
\end{equation}

\begin{equation}
r_{t+1}=\rho r_{t}+(1-\rho)\nabla\hat{L}_{G_{t+1}}^{w_{i}}\odot\nabla\hat{L}_{G_{t+1}}^{w_{i}},r_{0}=0,
\end{equation}
where $\odot$ denotes the element-wise product operator, $\tau$ is the general learning rate of the model, $\delta$ is a small constant, and $r_t$ represents the accumulated gradient term. $\rho$ is a balancing term to retain the previous time slices' gradient weight, while $(1 - \rho)$ represents the gradient weight for the current time slice. To further reduce the impact of local optima, we perform an element-wise product between the self-adaptive weighted aggregated time slices and random binary masks and then sum all the results. This serves as the final aggregated gradient for the $i$-th sliding window:

\begin{equation}
\nabla\hat{\mathcal{L}}^{w_{i}}=\sum_{k=t+1}^{t+l-1}M_{k}\nabla\hat{\mathcal{L}}^{\prime}{}_{G_{k}}^{w_{i}},
\end{equation}
where $M$ is a random binary mask matrix. In this manner, GradWATCH can randomly select a batch of time-slice gradients across the dynamic graph and aggregate them into a comprehensive loss gradient for the window.

\subsection{Inter-window parameter update module}
Between windows, the graph encoding parameters of the next sliding window $w_{i+1}$ are updated by combining the aggregated gradient $\nabla\hat{\mathcal{L}}^{w_{i}}$ of the previous sliding window $w_{i}$ with the instantaneous time slice gradient $\nabla\mathcal{L}_{G_{t+l}}^{w_{i}}$ of the last time slice in the current window:

\begin{equation}
\theta_{t+l-1}^{w_{i+1}}\longleftarrow(\nabla\mathcal{L}_{G_{t+l-1}}^{w_{i}})^{\tau}+(\nabla\hat{\mathcal{L}}^{w_{i}})^{\tau}+\theta_{t+l-1}^{w_{i+1}}.
\end{equation}

As the sliding window moves, GradWATCH's parameters are continuously updated and propagated. Ultimately, we obtain the representations of all account nodes in the MixTAG, denoted $Z=\{z_1, z_2, ... \}$.

\subsection{Account association detection module}

In the inference phase, we use the trained node representations to predict links for all potential candidate account connections, including both positive and negative links. Specifically, we compute the similarity of the embeddings of two nodes $v_i$ and $v_j$ in the candidate set to generate the probability of the link's existence:
\begin{equation}\begin{aligned}
P\left(A|Z\right) & =\prod_{i=1}^{n}\prod_{j=1}^{n}P\left(A_{ij}|z_{i},z_{j}\right), \\
P\left(A_{ij}=1|z_{i},z_{j}\right) & =Sigmoid\left(\theta<z_{i},z_{j}>\right),
\end{aligned}\end{equation}
where $P\left(A|Z\right)$ represents the overall probability distribution of account association edges between nodes. $A$ is an $n \times n$ matrix, where each element $A_{ij}$ indicates whether an edge exists between node $v_i$ and $v_j$. And $Z$ is the set of embeddings for all nodes. The function $P(\cdot)$ represents the probability of an edge existing between nodes $v_i$ and $v_j$, which can be either the cosine similarity function or the inner product function.

\section{Experiment}
In this section, we conduct comprehensive experiments of GradWATCH, addressing the following key questions:
\begin{itemize}
\item {RQ1: How are the experimental data, evaluation metrics, and parameters configured?}
\item {RQ2: How does GradWATCH perform compared to existing account tracking methods?}
\item {RQ3: What impact do different components of GradWATCH have on the experimental results?}
\item {RQ4: How robust is GradWATCH when labeled data is sparse? How sensitive is GradWATCH to parameter variations?}
\item {RQ5: How does GradWATCH compare with baseline methods in terms of computational efficiency?}
\end{itemize}
\subsection{Experimental Setup (RQ1)}
\subsubsection{Datasets}
The datasets consist of transactions and label information. We collect raw transaction records related to mixing services from the Ethereum blockchain explorer, Etherscan\footnote{https://etherscan.io/}, covering transactions associated with mixing service contracts for different denominations (0.1 ETH, 1 ETH, 10 ETH, and 100 ETH). The datasets include 526,472 transactions from “2019-12-16" to “2025-01-06". After processing, we obtain 150,315 purified transactions, with 57,064 for 0.1 ETH, 11,098 for 1 ETH, 11,080 for 10 ETH, and 71,073 for 100 ETH. For labelling, we adopt two heuristic methods inspired by previous research \cite{watching,bert4eth,shenmeng}: (1) transaction-based dataset. (2) ENS-based dataset. A detailed description is provided in Section \ref{sec:section3.1} Stage 1, and the details of the dataset are shown in Table \ref{tab:3}. 
% For dynamic graph construction, we employ different temporal slicing strategies for two datasets due to their distinct data distributions. The Transaction-based dataset uses equal-time slicing, while the ENS-based dataset adopts equal-count slicing to ensure that each time slice contains sufficient labeled account associations.
The density \( D \) of MixTAG is defined as:
$D = \frac{|\mathcal{E}|}{n(n-1)},$
where \( |\mathcal{E}| \) denotes the number of edges and \( n \) is the total number of nodes. The density value \( D \) is within the interval \([0, 1]\). A density approaching \( 0 \) indicates a highly sparse graph, whereas a value approaching \( 1 \) implies a densely connected structure. Furthermore, “Avg. graph density" in the Table \ref{tab:3} represents the average density of all time slices. “Avg. density increase" indicates the degree of improvement in the average graph density after using our graph structure, which is 21.9 times and 2506 times, respectively. 

It is worth noting that although the two datasets have the same number of account nodes and transaction edges, their average graph densities and corresponding density increments differ significantly. These differences mainly stem from differences in the number of account association labels and the dynamic graph construction strategies adopted.
Specifically, the Transaction-based dataset contains a relatively larger number of account association labels that are more widely dispersed over time, and therefore adopts an equal-time slicing strategy with fixed-length intervals. In contrast, the ENS-based dataset includes far fewer labeled account associations with a more uneven temporal distribution, and thus employs an equal-count slicing strategy to ensure sufficient supervision signals within each time slice. As a result, equal-time slicing may produce a large number of sparse slices in practice. These low-density slices effectively dilute the average relative density change, such that even with a large number of account association labels, the overall average density increment remains limited. In contrast, equal-count slicing leads to a more stable edge distribution and a higher average graph density. When association edges are introduced, the relative density fluctuations across different slices become more pronounced, leading to a larger average density increment.

% \begin{table}[htbp]
%   \fontsize{7}{11.5}\selectfont % 设置字体大小为 5 磅，行间距为 10 磅 \baselineskip
%   \setlength{\tabcolsep}{0.8pt} % 设置列间距为 6pt
%   \caption{Dataset information for the six types of accounts.}
%   \label{tab:3}
%   \begin{tabular}{|c|c|c|}
%     \hline
%     \textbf{Dataset}&\multicolumn{1}{c|}{\textbf{Heuristic methods}}&\multicolumn{1}{c|}{\textbf{Heuristic methods}} \\
%     % \cline{2-3} 
%     \textbf{information} & \textbf{based on transaction}&\multicolumn{1}{c|}{\textbf{based on ENS}}\\
%     % 
%     \hline
%     \textbf{Num. of positive samples} & 16,803 & 155\\
%     \hline
%     \textbf{Ave. Num. of positive samples in time slice} & 336.06 & 155\\
%     \hline
%     \textbf{Graph} & 50 & 310\\
%     \hline
%     \textbf{Ave. Num. of nodes} & 107,000 & 84.62 \\
%     \hline
%     \textbf{Ave. Num. of edges} & 7,016.8 & 178.34 \\
%     \hline
%   \end{tabular}
% \end{table}

\begin{table}[!t]
  \fontsize{9.5}{12}\selectfont % 设置字体大小为 8 磅，行间距为 12 磅
  \setlength{\tabcolsep}{3pt} % 设置列间距为 8pt
  \caption{Dataset information for account tracking task.}
  \label{tab:3}
  \begin{tabular}{c c c} % 不需要在两边加上竖线
 \toprule
 \midrule
    % \textbf{Dataset}&\multicolumn{1}{c}{\textbf{Heuristic methods}}&\multicolumn{1}{c}{\textbf{Heuristic methods}} \\
    % % \cline{2-3} 
    % \textbf{information} & \textbf{based on transaction}&\multicolumn{1}{c}{\textbf{based on ENS}}\\
    Dataset& Transaction-based &ENS-based \\
    % \textbf{information} & \textbf{based on transaction}&\multicolumn{1}{c}{\textbf{based on ENS}}\\
    \hline
    Num. of labels & 16,805 & 140 \\
    Num. of time slices & 100 & 20 \\
    % \hline
    % \hline
    % \hline
    Num. of nodes & 107,000 & 107,000 \\
    % \hline
   Avg. num. of transaction edges & 3,508.14 & 17,540.7 \\
    Avg. account associated edges & 168.05 & 7.00 \\
   Avg. graph density & 3.21$\times 10^{-7}$ & 1.53$\times 10^{-6}$ \\
   Avg. density increment & 21.9 & 2,506 \\
\midrule
\toprule
  \end{tabular}
\end{table}

\subsubsection{Evaluation metrics}
To evaluate GradWATCH's performance in account tracking tasks, we use the following seven metrics. The area under the ROC curve (AUC) measures the model's ability to distinguish between positive and negative samples. Then, we use five common metrics: Recall, Precision, $F_1$, FPR, and FNR. 
% A higher precision indicates a lower likelihood of misclassifying negative samples as positive, while a higher recall reflects the model's stronger ability to identify true positive samples. The F1-score, as the harmonic mean of precision and recall, captures the overall balance of the model between these two aspects.
Additionally, we use the mean reciprocal rank (MRR) to assess the model's ranking performance on positive samples. MRR calculates the average reciprocal rank of the first correct prediction in all queries, providing information on the effectiveness of the model's ranking, which is defined as follows:
$MRR=\frac{1}{|Q|}\sum_{i=1}^{|Q|}\frac{1}{rank_i}$,
where $|Q|$ is the total number of queries and ${rank_i}$ represents the position of the first correctly predicted positive sample in the ranked list for query $i$.
\subsubsection{Parameter setting}
GradWATCH is implemented based on PyTorch and the DGL library. When transforming transaction-level features into account-level features, we set the dimension of the generated node embeddings to 544. Specifically, the transaction type, transaction value, and transaction time are embedded in 128 dimensions. The noise term has 256 dimensions, the position embedding has 16 dimensions, and the edge features projected into the node space contribute 14 dimensions. During graph encoding, we employ a two-layer edge-aware graph encoder with a hidden-layer dimension of 256, a final node embedding dimension of 64, and a dropout rate of 0.1. The initial general learning rate is 0.02 for optimization, while the meta-learning rate is 0.008. The empirically determined optimal window sizes are 15 for the transaction-based heuristic dataset and 5 for the ENS-based heuristic dataset.

\subsection{Comparison Methods}
To thoroughly evaluate the performance of our proposed method, we categorize the baselines into three groups.
\textbf{(1) Traditional machine learning methods.} This includes feature-driven models such as Logistic Regression (LR) \cite{lr} and Random Forest (RF) \cite{rf}, as well as sequence modeling techniques like Recurrent Neural Networks (RNN) \cite{rnn} and Long Short-Term Memory (LSTM) \cite{lstm}, which are used to capture temporal patterns.
\textbf{(2) Graph learning methods.} This includes 1) graph embedding methods such as DeepWalk \cite{deepwalk} and Node2Vec \cite{node2vec}, which model account relationships through random walks; 2) static graph neural networks including four widely-used architectures: GCN \cite{gcn}, GAT \cite{gat}, GIN \cite{gin}, and GraphSAGE \cite{graphsage}; and 3) dynamic graph learning models such as DGNN \cite{dgnn}, EvolveGCN \cite{evolveGCN}, DyGraph2Vec \cite{dyngraph2vec}, and WinGNN \cite{wingnn}, which focus on modeling the temporal evolution of account interactions.
\textbf{(3) Recent advanced methods.} This addresses the account tracing task without relying on heuristic rules, such as Bert4eth \cite{bert4eth} and Mixbroker \cite{shenmeng}. 
In addition, we generate 212-dimensional manual node features for each account from the purified transactions. These features serve as the initial account features for the baseline methods. 
% They are also used in the ablation study to replace the initial node features when the transaction-to-account mapping module is removed. 
More details are presented in Table \ref{tab:transaction_features}.

\begin{table*}[!t]
    \centering
    \fontsize{10}{12}\selectfont
    \caption{Manual feature explanation for account (d: deposit, w: withdrawal, a: all, t: total, x: stage index).}
    \label{tab:transaction_features}
    \resizebox{\textwidth}{!}{ % 将表格宽度强制设为页面宽度
        \begin{tabular}{lccp{8cm}} % 这里的 p 宽度可以设大一点，resizebox 会处理剩下的
        \toprule
        \midrule
        Feature Category & Num. of features & Feature Names and Descriptions \\
        \midrule
        \multirow{5}{*}{Numeric Features (32Dim)} 
            & 3 * 5 & \{0.1, 1, 10, 100, all\}\_num\_\{d, w, a\}: Number of transactions for different purposes and different amounts. \\
            & 2 * 3 & \{avg, std\}\_num\_\{d, w, a\}: Avg and std of the number of transactions for different purposes. \\
            & 2 * 2 & \{value, avg\_value\}\_\{d, w\}: Total and avg transaction values for different purposes.\\
            & 2 & d/w\_value, avg\_d/w\_value\_ratio: Deposit/Withdrawal transaction volumes and averages. \\
            & 5 & d/w\_num\_\{0.1, 1, 10, 100, a\}: Ratio of deposit and withdrawal transactions for different purposes. \\
        \midrule
        \multirow{1}{*}{Gas Fee Features (45Dim)}
            & 3 * 3 * 5 & gf\_\{min, max, avg\}\_\{d, w, a\}\_\{0.1, 1, 10, 100, a\}: Max, min, and all gas fees by purpose and amount. \\
        \midrule
        \multirow{2}{*}{Time Features (90Dim)}
            & 2 * 3 * 5 & time\_\{e, l\}\_\{d, w, a\}\_\{0.1, 1, 10, 100, a\}: Earliest and latest transaction times by purpose and amount. \\
            & 4 * 3 * 5 & timegap\_\{min, max, avg, t\}\_\{d, w, a\}\_\{0.1, 1, 10, 100, a\}: Time gaps by purpose and amount. \\
        \midrule
        \multirow{3}{*}{Cross Features (54Dim)}
            & 3 * 5 & tps\_\{d, w, a\}\_\{0.1, 1, 10, 100, a\}: Transaction frequency by purpose and amount. \\
            & 3 & gas\_eff\_\{d, w, a\}: Gas fee to transaction amount ratio by purpose. \\
            & 12 * 3 & bin\_x\_\{d, w, a\}: Transaction counts by purpose and time period ($x \in [1, 12]$).\\
        \midrule
        \bottomrule
    \end{tabular}
 }
\end{table*}

% \begin{table}[ht]
% \centering
% \caption{Manual node features for each account.}
% \label{tab:transaction_features}
% \begin{tabular}{ll}
% \toprule
% \textbf{Feature} & \textbf{Description} \\
% \midrule
% min\_time\_gap & Minimum time interval for all transactions. \\
% min\_time\_gap\_d & Minimum time interval for all deposit transactions. \\
% avg\_time\_gap & Average time interval for all transactions. \\
% max\_time\_gap & Maximum time interval for all transactions. \\
% late\_time & Time of occurrence of the last transaction. \\
% avg\_time\_gap\_d & Average time interval for all deposit transactions. \\
% late\_time\_d & Time of occurrence of the last deposit transaction. \\
% early\_time & Time of occurrence of the first transaction. \\
% max\_time\_gap\_d & Maximum time interval for all deposit transactions. \\
% min\_time\_gap\_w & Minimum time interval for all withdrawal transactions. \\
% \bottomrule
% \end{tabular}
% \end{table}

\newcolumntype{Y}{>{\centering\arraybackslash}X}
\begin{table*}[!t]
\small
\caption{Performance evaluation of GradWATCH and baselines under standard deviation across multiple metrics on transaction-based heuristic dataset. The \textbf{best} and the \underline{second} are marked. $(\uparrow)$ / $(\downarrow)$ indicates the direction where the metric is better.}
\label{tab:4}
\renewcommand{\arraystretch}{1.2}
\begin{tabularx}{\linewidth}{l *{7}{Y}}
  \toprule
  \midrule
  \multirow{2}{*}{Methods} 
  & \multicolumn{7}{c}{Transaction-based heuristic dataset} \\
  \cmidrule{2-8}
  & AUC $(\uparrow)$ & Recall $(\uparrow)$ & Precision $(\uparrow)$ & $F_1$ $(\uparrow)$&MRR $(\uparrow)$ & FPR $(\downarrow)$ & FNR $(\downarrow)$ \\
  \midrule
  LR \cite{lr}        &$82.97\pm1.36$ &$58.21\pm2.02$ & $81.72\pm2.05$& $67.98\pm1.86$ &$43.04\pm1.60$ &$18.04\pm1.30$ & $41.79\pm2.02$\\ 
  RF  \cite{rf}       &$\underline{83.98\pm0.47}$ & $80.80\pm1.37$& $83.81\pm1.31$ & $\underline{82.32\pm0.61}$ &$47.57\pm1.68$ & $17.58\pm1.68$ & $\underline{9.20\pm1.37}$\\
  RNN   \cite{rnn}     & $52.95\pm1.77$ & $30.28\pm4.67$ & $52.94\pm2.16$ & $36.50\pm9.42$ &$62.15\pm2.67$ & $27.81\pm1.59$ & $69.72\pm4.67$  \\
  LSTM  \cite{lstm}     &$51.75\pm1.01$ &$49.65\pm9.13$ &$52.49\pm2.43$ &$48.83\pm9.52$ &$50.00\pm1.09$ &$46.07\pm9.56$ & $50.35\pm9.13$\\ \hline
  DeepWalk \cite{deepwalk}  &$60.27\pm0.73$ &$59.98\pm0.84$ &$79.75\pm4.58$ & $68.47\pm1.26$& $52.30\pm3.03$ & $18.85\pm1.09$ & $33.47\pm0.83$ \\
  Node2Vec \cite{node2vec}  &$62.91\pm2.37$ &$61.32\pm0.81$ &$81.85\pm3.83$ &$70.12\pm1.36$ &$58.21\pm1.64$ &$19.10\pm0.44$ &$39.17\pm0.81$ \\
  GCN \cite{gcn}       &$63.49\pm1.59$ &$76.83\pm1.78$ &$53.19\pm1.90$ &$62.83\pm1.32$ &$59.84\pm1.02$ & $67.71\pm4.13$ & $23.16\pm1.78$ \\
  GAT  \cite{gat}      &$70.03\pm3.42$ &$83.35\pm5.35$ &$54.35\pm1.70$ &$66.67\pm2.69$ &$60.07\pm2.17$ &$72.44\pm3.49$ &$13.65\pm5.35$\\
  GIN   \cite{gin}     &$66.81\pm1.52$ &$26.24\pm1.07$ &$\underline{92.49\pm2.37}$ &$40.85\pm1.15$ &$64.25\pm1.01$ &$\underline{17.57\pm0.82}$ &$73.73\pm1.07$\\
  GraphSAGE \cite{graphsage} &$67.90\pm2.04$ &$79.91\pm5.42$ &$56.55\pm2.38$ &$66.06\pm1.31$ &$69.10\pm1.02$ & $61.91\pm8.90$ & $20.09\pm5.42$ \\
  \hdashline
  DGNN \cite{dgnn} &$79.94\pm3.79$ &$60.00\pm2.13$ &$52.03\pm1.90$ &$55.01\pm1.89$ &$54.77\pm2.17$ &$35.69\pm2.91$ & $41.00\pm2.14$\\
  EvolveGCN-O \cite{evolveGCN} &$61.42\pm9.82$ &$70.62\pm5.35$ &$72.55\pm8.05$ &$64.12\pm3.89$ &$62.60\pm0.21$ & $47.78\pm9.28$ & $29.38\pm2.35$ \\
  EvolveGCN-H \cite{evolveGCN} &$80.74\pm1.19$ &$83.54\pm4.73$ &$78.08\pm3.36$ &$81.79\pm0.83$ &$62.60\pm0.11$ &$25.06\pm6.40$ & $13.46\pm4.73$\\
  Dygraph2vec \cite{dyngraph2vec} &$83.65\pm1.07$ &$60.90\pm3.46$ &$91.13\pm1.41$ &$71.86\pm1.57$ &$\underline{75.99\pm0.84}$ &$17.58\pm3.71$ & $39.10\pm2.33$\\
  WinGNN \cite{wingnn} &$80.36\pm1.37$ &$74.51\pm3.11$ &$76.56\pm1.85$ &$75.52\pm2.33$ &$73.91\pm0.93$ &$22.80\pm3.71$ & $25.49\pm2.33$\\
   \hline
  Bert4ETH \cite{bert4eth}   &$63.20\pm1.07$ &$\boldsymbol{89.13\pm1.07}$ &$50.93\pm1.07$ &$64.82\pm1.07$ &$32.51\pm1.07$ &$84.57\pm1.07$ &$10.87\pm1.07$\\
  Mixbroker \cite{shenmeng} & $56.37\pm0.56$ & $38.05\pm0.64$ & $60.08\pm1.04$ & $46.59\pm0.62$ & $53.27 \pm 2.34$ &$25.31\pm1.09$ &$61.95\pm0.64$ \\
  GradWATCH & $\boldsymbol{97.07\pm0.76}$ & $\underline{84.96\pm2.32}$ & $\boldsymbol{94.91\pm3.06}$ & $\boldsymbol{89.63\pm2.17}$ &  $\boldsymbol{91.21\pm2.95}$ & $\boldsymbol{17.33\pm3.00}$ & $\boldsymbol{7.75\pm1.17}$\\
  \hline
  \%Improve.   &$13.09\%$ &$-4.17\%$ &$2.42\%$ &$7.31\%$ &$15.22\%$ &$0.24\%$ & $1.45\%$ \\
  \midrule
  \bottomrule
\end{tabularx}
\end{table*}
\subsection{Analysis of experimental results (RQ2)}
In the account tracking task, we evaluate GradWATCH on two heuristic datasets of varying scales (transaction-based and ENS-based) to validate its performance across different graph densities and label counts. As shown in Tables \ref{tab:4} and \ref{tab:5}, GradWATCH achieves state-of-the-art results across all six key metrics.

% In the transaction-based heuristic dataset (\textbf{\textit{high graph density, abundant labels}}), GradWATCH achieves an AUC of 97.07\%, significantly outperforming all baseline methods. It also achieves an $F_1$ of 89.63\%, demonstrating a well-balanced performance between precision (94.91\%) and recall (84.96\%).
On the transaction-based heuristic dataset \textbf{\textit{(high graph density, abundant labels)}}, GradWATCH achieves an AUC of 97.07\%, significantly outperforming all baseline methods. It also achieves an F1 of 89.63\%, demonstrating a well-balanced performance between Precision (94.91\%) and Recall (84.96\%). Experimental results show that traditional machine learning methods and static graph learning approaches tend to excel in either Precision or Recall, but struggle to achieve a favorable balance between the two. Existing account tracking methods, such as Bert4eth, achieve high Recall but suffer from low Precision. This limitation stems from their behavior-similarity-based retrieval strategies, which generate overly broad candidate sets in highly homogeneous transaction environments, resulting in a large number of false-positive matches. In contrast, GradWATCH enforces both structural and temporal consistency within a dynamic graph, predicting control links only when sufficient and consistent evidence is observed across time windows. This design effectively reduces false positives and improves detection reliability. As a result, our method achieves an MRR of 91.21\%, representing a 15.22\% improvement over the best competing method.
% Experimental results show that traditional machine learning methods and static graph learning approaches tend to excel in either precision or recall, but not both. 
% Dynamic graph methods show stable performance but lack comprehensive effectiveness. Existing account tracking methods, such as Bert4eth, achieve high recall but suffer from low precision, as their behavior-similarity–based retrieval strategy produces overly broad candidate sets in highly homogeneous transaction environments, thereby introducing a large number of false positive matches. In contrast, GradWATCH achieves an MRR of 91.21\%, marking a 15.22\% improvement over comparative solutions, validating its effectiveness in priority ranking tasks.

% In the ENS-based heuristic dataset (\textbf{\textit{low graph density, sparse labels}}), as the graph complexity increased and the availability of the labels decreased, all methods experienced performance degradation. However, as shown in Table \ref{tab:5}, GradWATCH consistently led in all five metrics, particularly in critical metrics such as AUC (84.26\%) and MRR (68.22\%), outperforming the second-best methods by 10.95\% and 1.62\%, respectively. This confirms the model's strong robustness against sparse graph data and label scarcity.

On the ENS-based heuristic dataset (\textbf{\textit{low graph density, sparse labels}}), all methods degrade in performance as graph sparsity increases and label availability decreases. Nevertheless, as shown in Table \ref{tab:5}, GradWATCH consistently outperforms all baselines across all five evaluation metrics, particularly in critical measures such as AUC (84.26\%) and MRR (68.22\%), surpassing the second-best methods by 10.95\% and 1.62\%, respectively. These results suggest that GradWATCH mitigates graph sparsity through its graph construction strategy, while the edge-aware window gradient propagation mechanism provides essential temporal reasoning signals, enabling robust performance even under sparse dynamic graphs and limited supervision. We further analyze the reasons why GradWATCH does not achieve the highest Recall on either dataset. First, this outcome reflects the inherent trade-off between Precision and Recall. Second, due to the long-tail distribution of controlled account relationships, the sliding-window modeling approach may fail to aggregate sufficient contextual information when related samples span multiple time windows.

In summary, although our method does not guarantee optimal performance across all seven evaluation metrics, it delivers more accurate and stable account association detection overall, with exceptional robustness observed particularly in challenging conditions involving sparse transactions and highly imbalanced labels.
\newcolumntype{Y}{>{\centering\arraybackslash}X}
\begin{table*}[!!htbp]
\small
\caption{Performance evaluation of GradWATCH and baselines under standard deviation across multiple metrics on ENS-based heuristic dataset. The \textbf{best} and the \underline{second} are marked. $(\uparrow)$ / $(\downarrow)$ indicates the direction where the metric is better.}
\label{tab:5}
\renewcommand{\arraystretch}{1.2}
\begin{tabularx}{\linewidth}{l *{7}{Y}}
  \toprule
  \midrule
  \multirow{2}{*}{Methods} 
  & \multicolumn{7}{c}{ENS-based heuristic dataset} \\
  \cmidrule{2-8}
  & AUC $(\uparrow)$ & Recall $(\uparrow)$ & Precision $(\uparrow)$ & $F_1$ $(\uparrow)$&MRR $(\uparrow)$ & FPR $(\downarrow)$ & FNR $(\downarrow)$ \\
  \midrule
  LR  \cite{lr}        & $58.89\pm1.97$ &$43.36\pm2.35$ &$\underline{63.07\pm2.90}$ &$49.98\pm1.82$ &$24.55\pm3.21$ &$24.55\pm3.21$ &$56.64\pm1.35$ \\ 
  RF  \cite{rf}       &$61.67\pm1.64$ &$56.67\pm2.00$ &$59.53\pm1.33$ &$58.06\pm1.88$ &$22.48\pm5.44$ &$22.48\pm5.44$ &$\underline{15.33\pm1.00}$\\
  RNN  \cite{rnn}      &$48.89\pm2.89$ &$58.83\pm1.73$ &$53.36\pm2.97$ &$52.10\pm1.82$ &$\underline{52.76\pm2.89}$ &$62.76\pm2.89$ & $41.17\pm1.73$ \\
  LSTM  \cite{lstm}     &$40.00\pm3.89$ &$24.19\pm3.79$ &$23.62\pm2.96$ &$21.72\pm2.98$ &$50.00\pm3.24$ &$39.73\pm2.42$ &$75.81\pm2.79$\\ \hline
  % DeepWalk   & & & & & \\
  % Node2Vec   &$62.91\pm2.37$ &$10.83\pm0.81$ &$61.05\pm3.37$ &$61.05\pm3.37$ &$61.05\pm3.37$ \\
  GCN \cite{gcn}       &$63.14\pm2.24$ &$27.31\pm2.17$ &$53.05\pm4.29$ &$34.07\pm3.84$ &$26.38\pm2.98$ &$29.5\pm3.50$ &$72.69\pm2.17$\\
  GAT \cite{gat}       &$66.18\pm1.27$ &$40.00\pm3.39$ &$62.31\pm2.89$ &$48.72\pm1.31$ &$27.09\pm3.27$ &$28.48\pm5.53$ &$60.00\pm3.39$ \\
  GIN  \cite{gin}      &$65.60\pm8.25$ &$27.14\pm8.25$ &$55.14\pm3.61$ &$34.03\pm2.75$ & $25.46\pm3.63$ &$\underline{23.15\pm11.37}$ &$7.86\pm8.25$ \\
  \hdashline
  DGNN \cite{dgnn} &$\underline{73.31\pm1.85}$ &$60.10\pm2.12$ &$52.03\pm2.01$ &$55.01\pm1.89$ &$42.96\pm3.31$ &$55.69\pm3.77$ & $40.00\pm2.12$\\
  GraphSAGE \cite{graphsage} &$54.51\pm2.71$ &$\underline{79.73\pm5.73}$ &$50.54\pm6.06$ &$61.01\pm6.52$ & $22.66\pm7.10$ & $77.14\pm8.66$ & $20.27\pm5.73$\\
  EvolveGCN-O \cite{evolveGCN} &$54.32\pm4.61$ &$68.43\pm9.56$ &$60.64\pm3.73$ & $53.79\pm3.57$& $24.03\pm0.01$ &$59.79\pm6.23$& $31.57\pm9.56$ \\
  EvolveGCN-H \cite{evolveGCN}  &$69.67\pm5.88$ &$76.42\pm9.36$ &$55.58\pm5.86$ &$\underline{64.37\pm5.27}$ &$24.03\pm0.14$ & $47.08\pm4.09$ & $15.58\pm9.36$\\
  Dygraph2vec \cite{dyngraph2vec} &$69.69\pm1.23$ &$71.00\pm0.11$ &$56.01\pm0.45$ &$62.62\pm0.23$ &$32.13\pm2.12$ & $25.01\pm4.08$ & $20.00\pm3.26$\\
  WinGNN \cite{wingnn} &$70.45\pm1.15$ &$71.85\pm0.15$ &$56.88\pm0.42$ &$63.50\pm0.25$ &$33.02\pm2.05$ & $24.15\pm3.95$ & $19.20\pm3.10$\\
  \hline
  Bert4ETH \cite{bert4eth}  &$59.66\pm1.27$ &$\boldsymbol{79.83\pm3.21}$ &$49.45\pm2.01$ &$61.06\pm1.68$ &$39.08\pm3.25$ &$\boldsymbol{21.34\pm1.07}$ &$21.71\pm2.46$\\
  Mixbroker \cite{shenmeng} & $50.93\pm9.21$ & $76.67\pm4.68$ & $51.69\pm5.61$ & $61.76\pm2.98$ & $20.83 \pm 1.94$ &$83.33\pm9.72$ &$15.33\pm4.68$ \\
  GradWATCH & $\boldsymbol{84.26\pm1.45}$ & $64.95\pm3.84$ & $\boldsymbol{71.77\pm2.91}$ & $\boldsymbol{68.22\pm2.32}$ & $\boldsymbol{54.38\pm5.26}$ &$26.76\pm11.75$& $\boldsymbol{14.45\pm9.93}$\\
  \hline
  \%Improve.  &$10.95\%$ &$-14.88\%$ &$8.70\%$ &$3.85\%$ &$1.62\%$ &$-5.42\%$ & $0.88\%$ \\
  \midrule
  \bottomrule
\end{tabularx}
\end{table*}

\begin{table}[!t]
\centering
\caption{Performance comparison of different model variants on transaction-based heuristic dataset.}
\fontsize{9}{12}\selectfont
\setlength{\tabcolsep}{4.5pt}
\begin{tabular}{lcccc}
\toprule
\midrule
Model Variant & AUC & Recall & Precision  & $F_1$ \\ 
\midrule
GradWATCH & \textbf{97.07} & \textbf{84.96} & \textbf{94.91} &\textbf{89.63}  \\ 
\hline
- \textit{w/o} Edge-Aware Graph Enc. &94.49  &78.86  &88.36 &80.34   \\
- \textit{w/o} T.-to-A. Map. & 92.67  &81.10  &83.97 &79.91      \\  
- \textit{w/o} MixTAG & 89.66 & 79.30 & 78.41 & 78.84 \\
% \hdashline
% \hdashline
- \textit{w/o} Intra-Win. Grad Com. &87.67  &78.24  &84.99 &81.48      \\
- \textit{w/o} Window &72.85  &78.78  &61.91 &69.33      \\  
% \hdashline
\midrule
\bottomrule
\end{tabular}
\label{tab:6}
\end{table}

\subsection{Ablation experiments (RQ3)}
\label{sec:section5.4}
\textit{RQ2} demonstrates the strong reasoning capabilities of our proposed GradWATCH for the account tracking task. To further validate the effectiveness of our well-designed components, we conduct ablation studies by sequentially removing or replacing key modules or mechanisms and evaluating their impact on overall performance.
In particular, \textit{“- w/o Edge-Aware Graph Enc."} denotes the ablation setting in which the edge-aware graph encoder is removed from the message passing process. As a result, the original transactional information carried by the edges is discarded, and the model resorts to a conventional GCN-based propagation mechanism. \textit{“-w/o T.-to-A. Map."} indicates the removal of the transaction-to-account mapping module, where randomly initialized node representations are used in place of the mapped embeddings. \textit{“- w/o MixTAG"} corresponds to the setting where only account-association edges are used to construct the homogeneous graph, while all other types of interactions are excluded. \textit{“- w/o Intra-Win. Grad Com."} variant removes the gradient computation and propagation between time slices within the window. \textit{“- w/o Window"} directly fixes the window size to a single evolution frame. It is important to note here that the experimental results of adjusting the window size are presented in the hyperparameter sensitivity analysis.

As shown in the ablation results in Table \ref{tab:6}, removing any module leads to performance degradation. Specifically, removing both the window and intra-window loss gradient computation and aggregation modules causes a 24.22\% decrease in AUC and a 21.30\% drop in $F_1$. When the window is preserved, but the gradient computation and propagation within the window are deleted, the AUC and $F_1$ are improved by 14.82\% and 12.15\%, respectively. However, this configuration still underperforms our model by 6.72\% to 9.92\% on all four metrics. Furthermore, eliminating MixTAG results in a 7.41\% and 16. 50\% decrease in Recall and Precision, respectively, demonstrating the necessity of incorporating a mixing transaction-driven account inference graph structure. Finally, after removing the edge-aware encoder and the transaction-account mapping module, the performance declined across all four metrics. It demonstrates that leveraging raw transaction information, heterogeneous edge-propagation and aggregation mechanisms, and extracting high-order account features from raw transactions are highly targeted and effective for enhancing account tracking performance under privacy-protection services.

\begin{figure}[h]
\centering
\includegraphics[width=7cm]{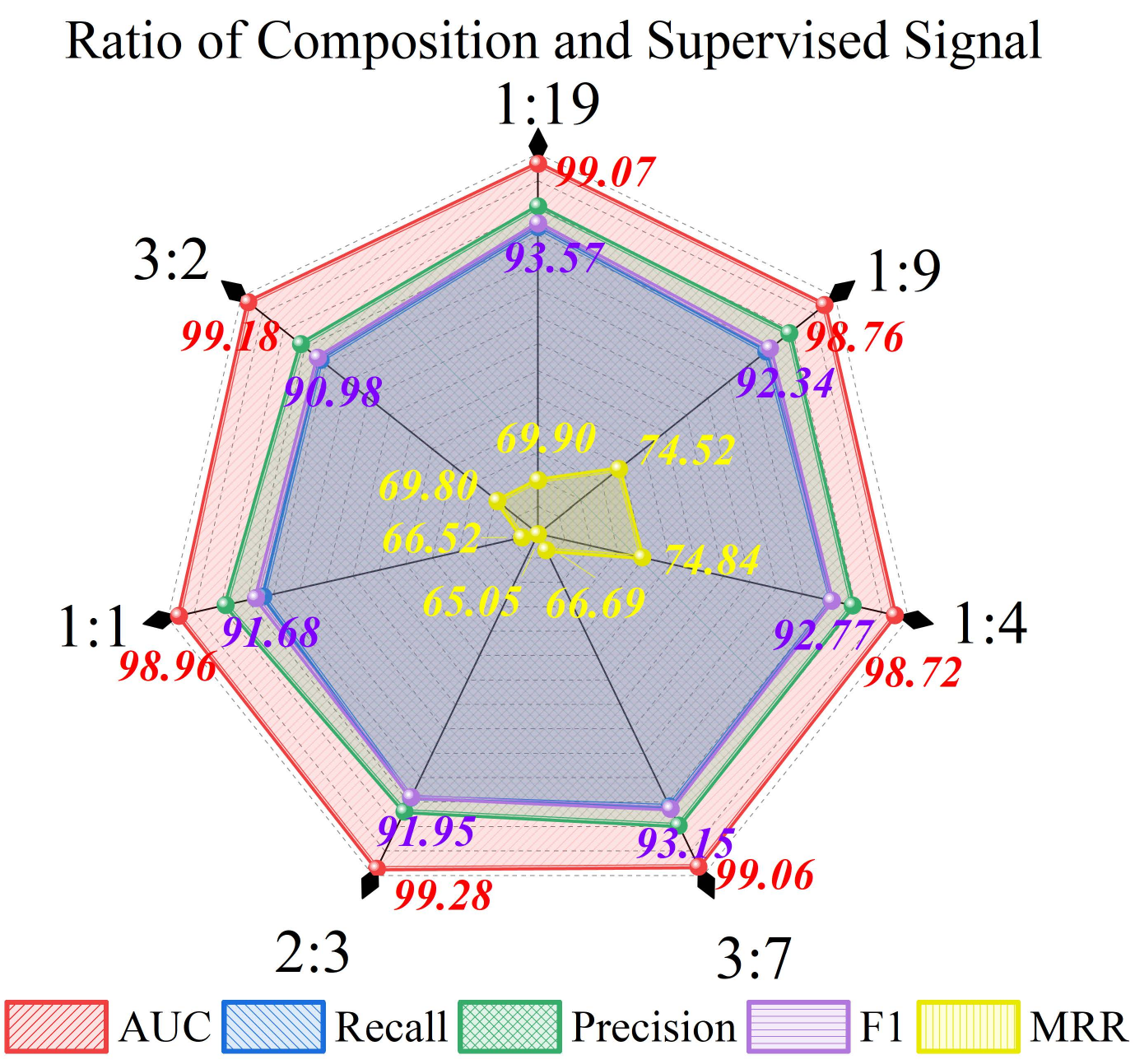}
\caption{Exploration of the optimal balance between labels for graph construction and supervision on the transaction-based heuristic dataset.}
\label{fig_7}
\end{figure}
\subsection{Model robustness (RQ4)}
After evaluating the model's overall performance, we investigated the influence of hyperparameter settings to validate the robustness of the proposed approach. Furthermore, we provide the following \textbf{Obs}ervations and interpretable insights.
% \subsubsection{\textbf{Optimizing Labels for the Best Balance of Composition and Supervision}}

\textbf{\textit{1) Obs. 1. An Optimal Balance Exists Between Labels for Graph Construction and Supervision.}}
Guided by structural learning theory \cite{StructuralStructural_Learning} and disentangled representation principles \cite{Disentangled_Representation}, we construct a dual-path label decoupling framework that enforces separation between \textit{“what"} (supervision signals) and \textit{“how"} (building graph) in account tracing tasks.  Specifically, raw labels are bifurcated into: i) binary classification supervision signals, and ii) compositional elements for constructing inter-account constraint edges in the graph. As illustrated in Fig. \ref{fig_7}, empirical results demonstrate optimal model performance in a 1:4 composition-to-supervision ratio ($F_1$ = 92.77\%, MRR = 74.84\%), revealing a dynamic equilibrium mechanism between structured priors and supervisory objectives. This balance suggests that moderate relational constraints enhance the capacity for reasoning about complex transactional patterns, whereas excessive structural learning introduces task-irrelevant feature interference.

\begin{figure}[!t]
\centering
\includegraphics[width=7cm]{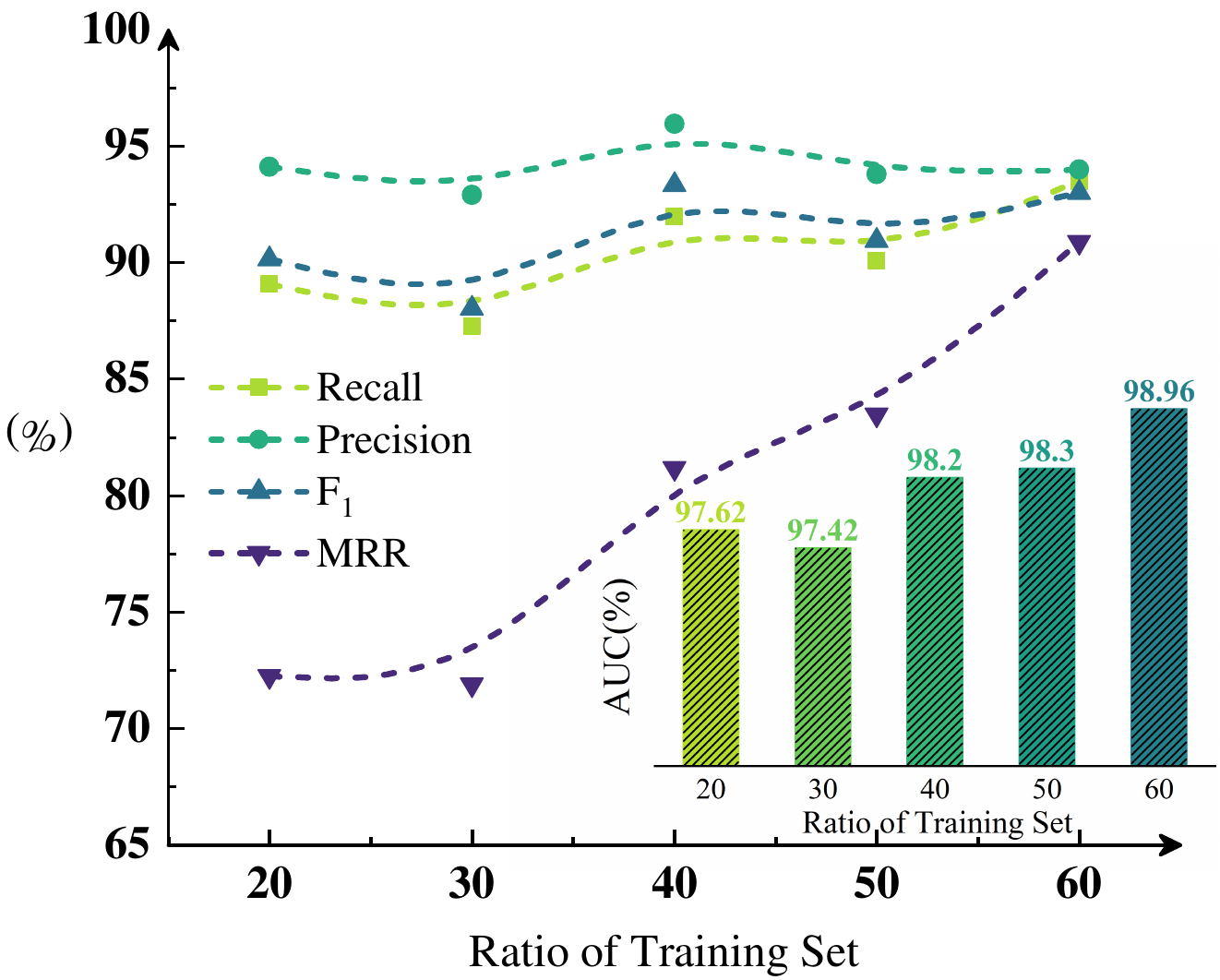}
\caption{Impact of training set proportion on performance on transaction-based heuristic dataset.}
\label{fig_9}
\end{figure}

\textbf{\textit{2) Obs. 2. Model Performance Remains Stable with Limited Labels.}}
In real-world scenarios, labels are often scarce. To simulate this condition, we adjust the proportions of training and testing samples within the entire dataset, and the experimental results are shown in Fig. \ref{fig_9}. Specifically, we vary the training set ratio from 20\% to 60\%, and correspondingly, the testing set ratio from 80\% to 40\%. The results show that core metrics such as AUC, Recall, Precision, and $F_1$ consistently remain around 90\%, demonstrating the model’s resilience. Notably, the MRR metric significantly improves when the training ratio exceeds 30\%. However, even with only 20\% of training data, the MRR score remains above 70\%. These results indicate that reducing the amount of labeled training data has only a marginal impact on overall performance, confirming the robustness of the proposed model.

% \subsubsection{Hyperparameter Sensitive Analysis}

\begin{figure}[!t]
    \centering
    \begin{subfigure}[b]{0.48\linewidth}
        \centering
        \includegraphics[width=\linewidth]{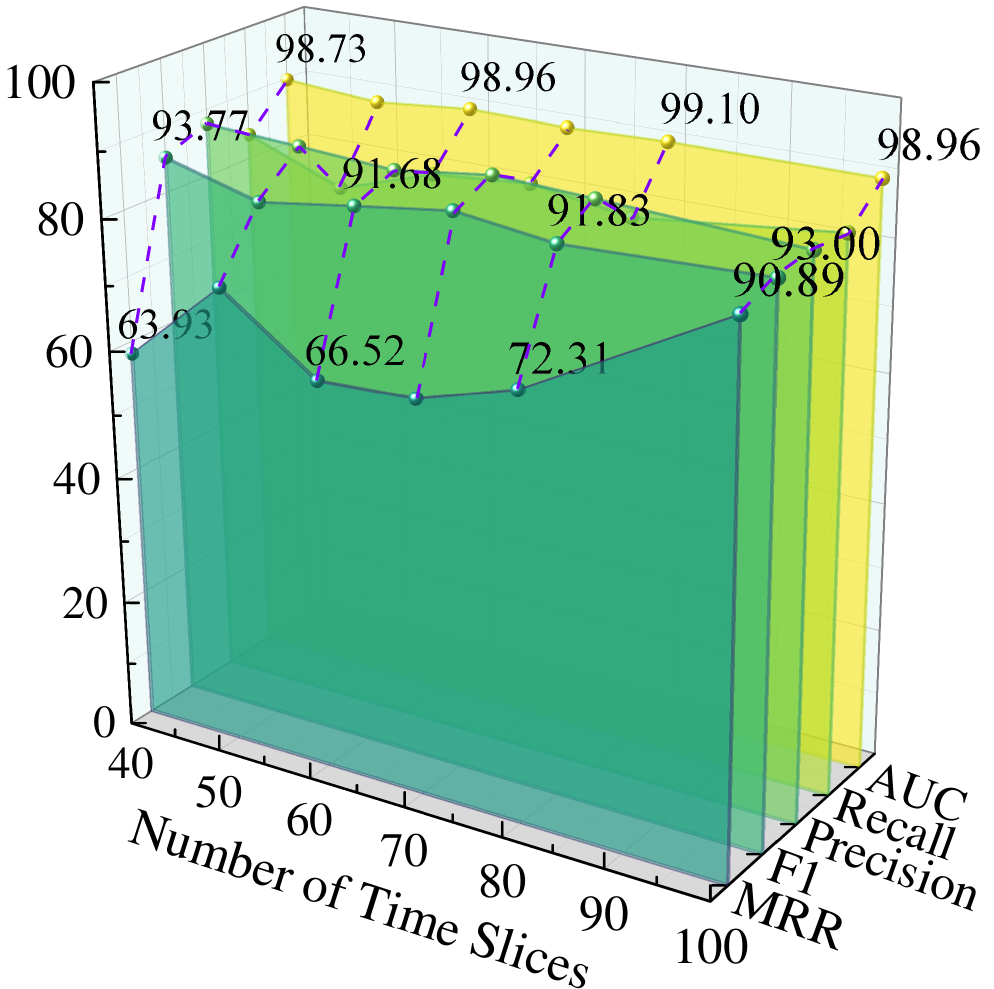}
        \caption{Effect of time slice.}
        \label{fig:subfig_a}
    \end{subfigure}
    \hspace{0pt} % 不加间距，避免溢出
    \begin{subfigure}[b]{0.48\linewidth}
        \centering
        \includegraphics[width=\linewidth]{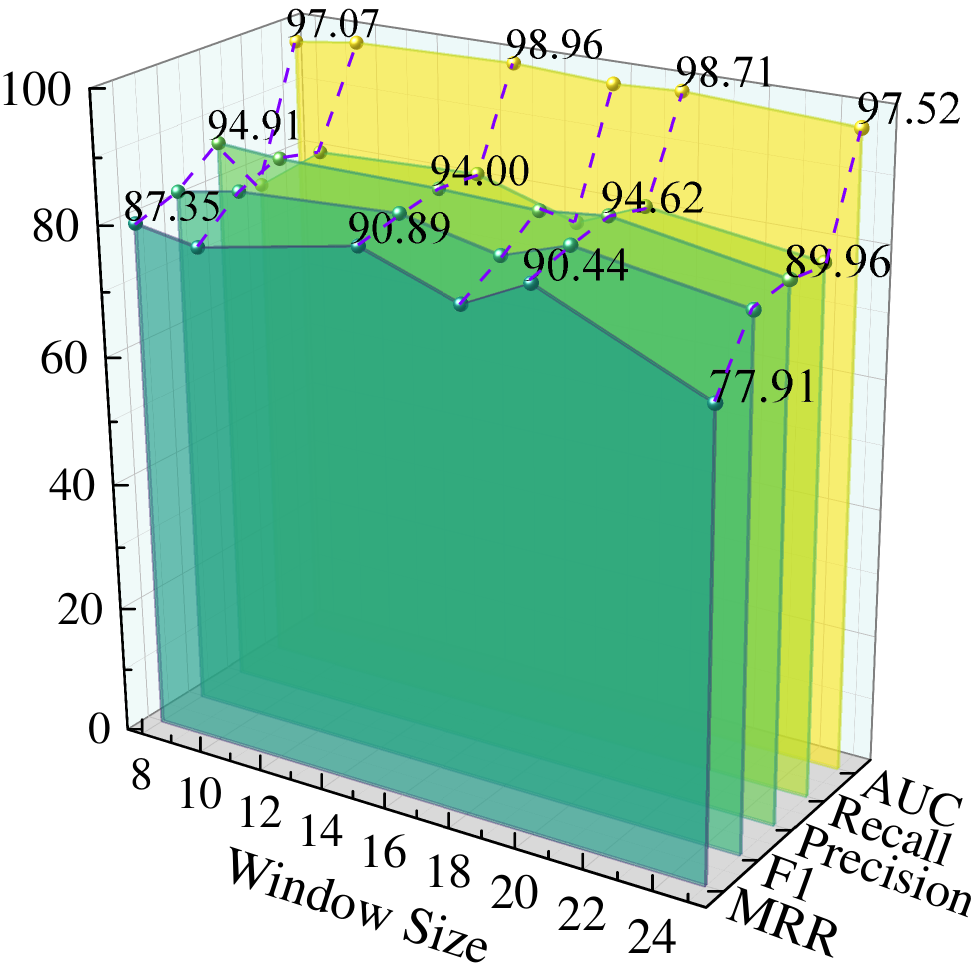}
        \caption{Effect of window size.}
        \label{fig:subfig_b}
    \end{subfigure}
    \caption{Time slice and window size are adjusted on the transaction-based heuristic dataset.}
    \label{fig_6}
\end{figure}

\textbf{\textit{3) Obs. 3. Model Robustness Under Hyperparameter Variations.}}
To investigate the sensitivity of our model to hyperparameters, we selected two critical hyperparameters—the number of time slices and the size of the sliding window—to evaluate their impact on performance. As shown in Fig. \ref{fig:subfig_a}, increasing the time slices from 40 to 100 consistently improves all evaluation metrics when tuning on the dataset based on mixing transactions. This indicates that 100 time slices is an optimal setting for this dataset. Moreover, the performance remains relatively stable throughout the tuning process, suggesting that our method is robust to changes in the number of time slices. 

Next, we fix the number of time slices to 100 and adjust the window size to explore the optimal configuration and its impact on experimental performance.  As shown in Fig. \ref{fig:subfig_b}, the model performance first improves and then declines as the window size increases from 8 to 25, reaching the highest performance when the window size is set to 15.  The result for the setting with the window size of 1 is reported under the ablation setting labeled \textit{“-w/o Window”} in \ref{sec:section5.4}. These results indicate that the window mechanism plays an important role in model performance, while the fluctuations during adjustment remain within a relatively narrow range.

% \begin{figure}[!t]
% \centering
% \includegraphics[width=7.5cm]{figures/time_slice_dif.pdf}
% \caption{The overview of our GradWATCH.}
% \label{fig_6}
% \end{figure}

\begin{table*}[t]
\centering
\caption{The results of average per-epoch runtime, peak memory usage, and inference throughput on the transaction-based heuristic dataset.}
\fontsize{9.5}{12}\selectfont
\begin{tabular}{lcccccc}
\toprule
\midrule
\multirow{4}{*}{Methods} & \multicolumn{3}{c}{Time(s) $(\downarrow)$}  & \multicolumn{2}{c}{Memory(GB) $(\downarrow)$} &Throughput (samples/s) $(\uparrow)$ \\
\cmidrule{2-7}
& Preprocessing Time & Training Time & Inference Time & Peak RAM & Peak GPU &Inference Throughput\\ 
% & Time & Time & Time &RAM &GPU &Throughput\\
\midrule
EvolveGCN-H & 110.31  & \textbf{42.36} & \textbf{0.73}  &29.37 &  7.36 &9208 \\ 
Dygraph2vec &110.31  &56.45  &0.78 &35.62 &8.37 &8618\\
WinGNN & 110.31  &44.23  &0.74 &29.92  & 6.29  &9084 \\  
Bert4ETH & 113.99 & 2,670.91 & 2154.48 & 7.01 & 16.41 &3.12\\
% \hdashline
% \hdashline
Mixbroker &100.32  &29.7  &0.57 &\textbf{1.06} & \textbf{0.29}  &\textbf{11793}  \\
\hline
GradWATCH &\textbf{90.89}  &62.08  &0.74 &30.93 & 6.50  &9084  \\  
% \hdashline
\midrule
\bottomrule
\end{tabular}
\label{tab:8}
\end{table*}

% \subsection{Study Case (RQ5)}
\subsection{Computational efficiency (RQ5)}
In this subsection, we investigate the computational efficiency of GradWATCH through both theoretical complexity analysis and experimental evaluation.

\textbf{\textit{1)	Theoretical Analysis of Time and Space Complexity.}}
The runtime of GradWATCH mainly consists of two components: (i) transaction-to-account mapping and (ii) edge-aware sliding-window encoding. Let $n$ be the number of account nodes and $T$ the number of time slices. We denote by $\bar{m}$ the average number of transaction edges per slice, $d_h$ the hidden dimension, and $d_1$ the dimension of the constructed account feature. The mapping step has a time complexity of $\mathcal{O}(N_{tx}\cdot d_1)$, where $N_{tx}$ is the number of transactions used to construct account features. During encoding, GradWATCH performs message passing over $T$ time slices, yielding $\mathcal{O}(T\cdot (\bar{m} \cdot d_h +n \cdot {d_h}^2))$, which is typically dominated by edge-wise aggregation in blockchain graphs and can be approximated as $\mathcal{O}(T\cdot \bar{m} \cdot d_h)$. Importantly, computation is performed only on observed transaction edges using sparse graph operators, avoiding dense pairwise interactions over all node pairs.

In terms of space complexity, GradWATCH stores $T$ temporal graphs and their features. The graph structure requires $\mathcal{O}(T\cdot \bar{m})$ memory for transaction edges, and $\mathcal{O}(T\cdot m_a)$ for association edges if included, where $m_a$ denotes the average number of association edges per time slice. Feature storage costs $\mathcal{O}(T\cdot n \cdot d_1)$ for account node features and $\mathcal{O}(T\cdot \bar{m} \cdot d_2)$ for edge features with dimension $d_2$. The model parameters scale as $\mathcal{O}(L\cdot {d_1}^2)$ for an $L$-layer encoder, while training-time activations introduce an additional $\mathcal{O}(L \cdot (n\cdot d_1+\bar{m}\cdot d_1))$ memory footprint. Overall, the dominant memory cost grows linearly with the number of stored snapshots and sparse edges.

\textbf{\textit{2) Efficiency Experimental Analysis.}} As shown in Table \ref{tab:8}, we report running time, peak memory usage (including RAM and GPU memory), and inference throughput of GradWATCH and several comparison methods with strong performance. The time-related metrics in the table include average processing time, training time, and inference time per epoch. Throughput is measured on the test set and reported in samples per second (samples/s). From the results, we can observe that MixBroker achieves high overall efficiency due to its simple modeling and method design, but its experimental performance is relatively poor. Bert4ETH maps each account address to a token in a dictionary, leading to an explosion in vocabulary size and, consequently, particularly high time overhead. Traditional dynamic graph processing methods require substantial time for feature engineering during preprocessing to construct optimal initial account features. In contrast, our end-to-end method incorporates task-specific optimization strategies for account association and achieves superior performance with only a tolerable increase in computational overhead.

\section{Conclusion}
With the rapid advancement of blockchain technology, address tracking has become a core research focus in compliance and anti-money laundering efforts. In this work, we propose GradWATCH, a novel gradient-aware dynamic window graph learning framework. Specifically, GradWATCH models transactions involving mixing services as dynamic, heterogeneous graphs. It employs a sliding window mechanism to propagate edge-aware graph encoder gradients across time slices within each window, and adaptively aggregates these instantaneous gradients to form cumulative gradients and window gradients. These gradients are then integrated to update model parameters, producing robust account representations. Finally, the method determines whether two addresses correspond to the same real-world entity by measuring similarity between learned representations. Experimental results across multiple datasets generated using heuristic rules and at different scales demonstrate that GradWATCH consistently achieves state-of-the-art performance. Future work will focus on enhancing the model's generalizability across cross-chain scenarios and on addressing challenges posed by evolving mixing protocols and nested money laundering strategies via cross-chain bridges.

\section{ACKNOWLEDGMENTS}
% This work is partially supported by the Beijing Natural Science Foundation (Z230001), the Beijing Advanced Innovation Center for Future Blockchain and Privacy Computing (GJJ-23-001, GJJ-23-002).
% 
This work was supported by the Beijing Natural Science Foundation (Grant No. Z230001), the National Natural Science Foundation of China (Grant No.62572029), the Beihang Ganwei Action Plan (Grant No. JKF-2025083360604), National Cyber Security-National Science and Technology Major Project under Grant (2025ZD1501002), Beijing Advanced Innovation Center for Future Blockchain and Privacy Computing, and China Scholarship Council (CSC).

\bibliography{references.bib} %bibfile_name

% \clearpage
\bibliographystyle{IEEEtran}
\begin{samepage}
% \setstretch{0.1}
% \vspace{-200pt} 
\begin{IEEEbiography}[{\includegraphics[width=1in,height=1.25in, clip,keepaspectratio]{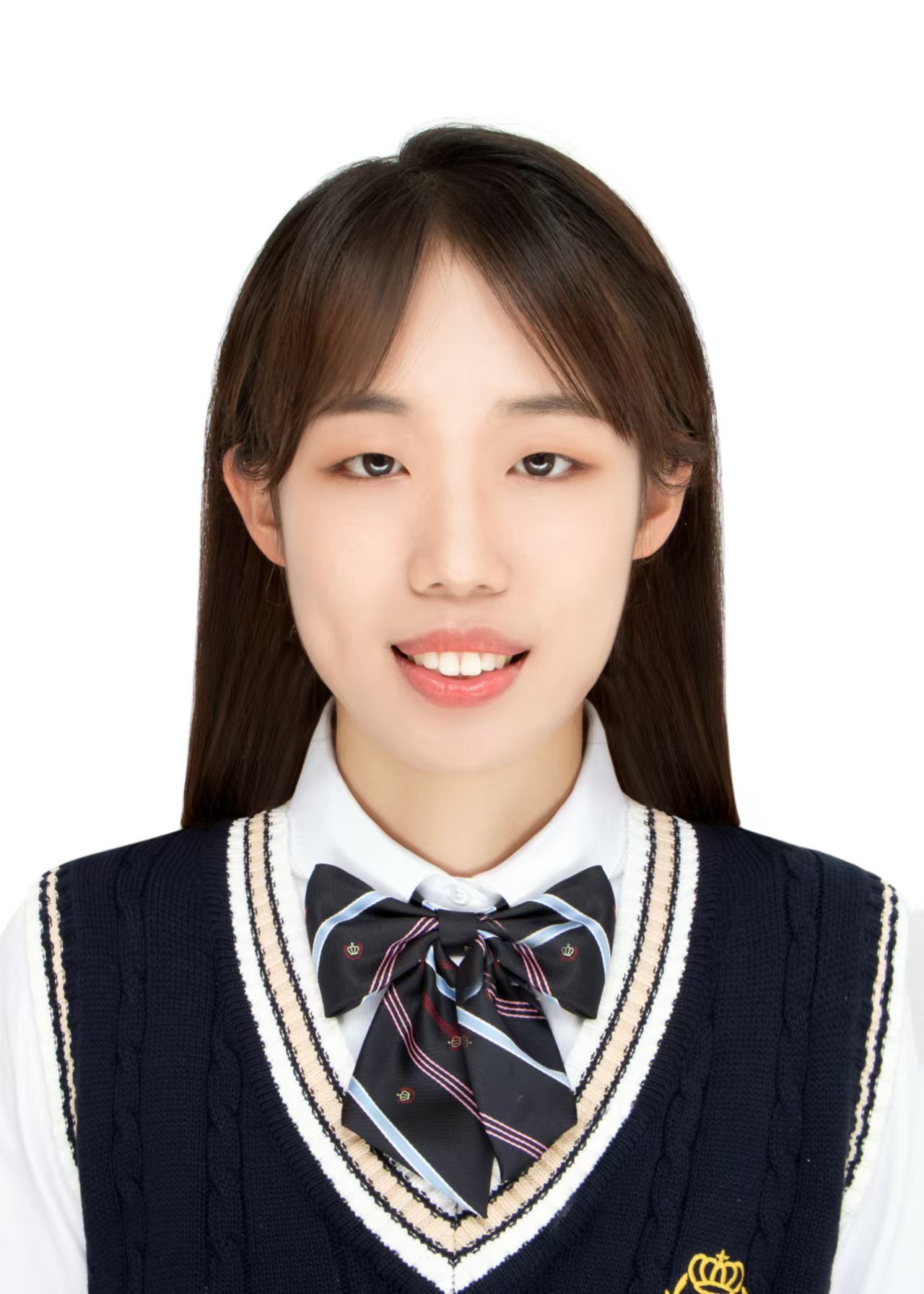}}]{Shuyi Miao} received the M.S. degree from the Department of Computer Science and Technology (College of Data Science), Taiyuan University of Technology, Taiyuan, China, in 2023. She is currently pursuing the Ph.D degree in the Institute of Artificial Intelligence at Beihang University, Beijing, China. Her research interests include blockchain regulation, graph machine learning, and knowledge graphs.
\end{IEEEbiography}\nopagebreak
% \vspace{-85pt} 
\begin{IEEEbiography}[{\includegraphics[width=1in,height=1.25in,clip,keepaspectratio]
{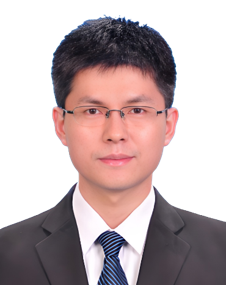}}]{Wangjie Qiu}
is currently an associate professor with the Beijing Advanced Innovation Center for Future Blockchain and Privacy Computing, Beihang University. He received his B.S. in 2008 and Ph.D. in 2013, both in mathematics from Beihang University. His research interests include Information security, blockchain, and privacy computing. He is also the deputy secretary-general of the Blockchain Committee at the China Institute of Communications. As a founding team member, he successfully developed ChainMaker, an advanced international blockchain technology system. He has been granted a number of invention patents in the fields of information security, blockchain, and privacy computing, and he won the first prize in the China National Technology Invention competition in 2014.
\end{IEEEbiography}\nopagebreak
% \vspace{-38pt} 
\begin{IEEEbiography}[{\includegraphics[width=1in,height=1.25in,clip,keepaspectratio]
{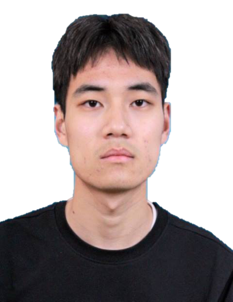}}]{Xiaofan Tu}
received his B.S. degree in 2024 from Wuhan University of Technology. He is currently pursuing the M.S. degree in the Institute of Artificial Intelligence at Beihang University, Beijing, China. His main research interests are blockchain, data mining, and graph learning.
\end{IEEEbiography}\nopagebreak
% \vspace{-55pt} 
\begin{IEEEbiography}[{\includegraphics[width=1in,height=1.25in,clip,keepaspectratio]
{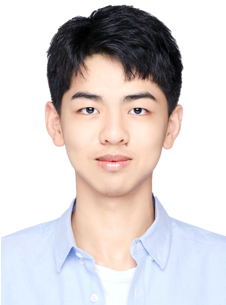}}]{Yunze Li}
received his B.S. degree in 2023 from Beijing University of Technology. He is currently pursuing the M.S. degree in the Institute of Artificial Intelligence at Beihang University, Beijing, China. His main research interests are blockchain, data mining, and graph learning.
\end{IEEEbiography}\nopagebreak
% \vspace{-55pt} 
\begin{IEEEbiography}[{\includegraphics[width=1in,height=1.25in,clip,keepaspectratio]
{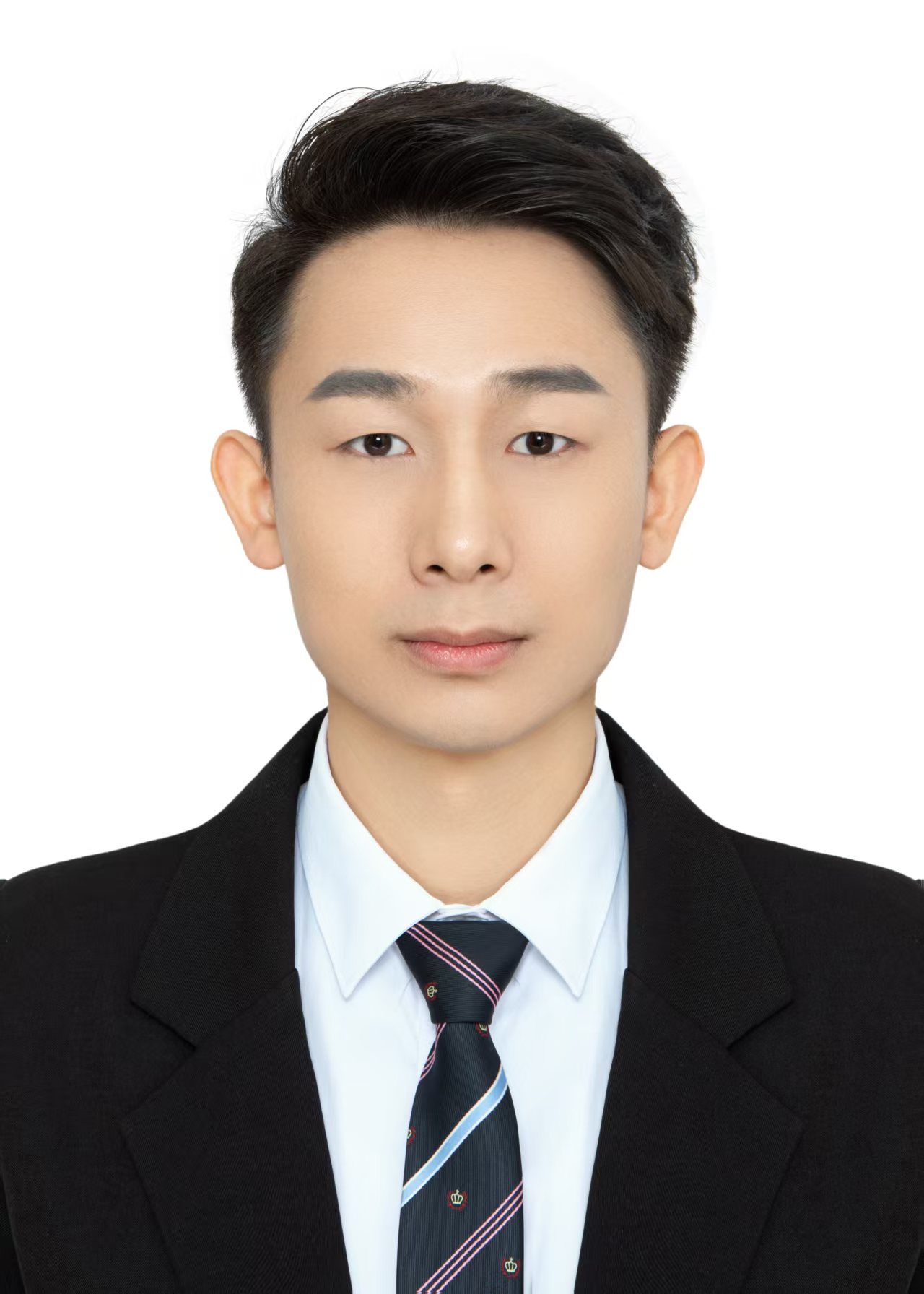}}]{Yongxin Wen} received the M.S. degree from the College of Information and Computer, Taiyuan University of Technology, Taiyuan, China in 2023. He received a B.S. degree in communication engineering from Northeastern University, Shenyang, China. He is currently working as an engineer at No.208 Research Institute of China Ordnance Industries, Beijing, China. His main research interests are channel prediction and backscatter communication. 
\end{IEEEbiography}\nopagebreak
% \vspace{-55pt} 
\begin{IEEEbiography}[{\includegraphics[width=1in,height=1.25in,clip,keepaspectratio]
{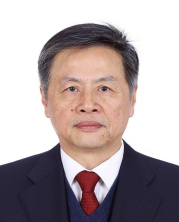}}]{Zhiming Zheng} received the Ph.D. degree in mathematics from the School of Mathematical Sciences, Peking University, Beijing, China, in 1987. He is currently a Professor with the Institute of Artificial Intelligence, Beihang University, Beijing, China. His research interests include refined intelligence, blockchain, and privacy computing. He is one of the initiators of Blockchain-ChainMaker. He is a member of the Chinese Academy of Sciences.
\end{IEEEbiography}\nopagebreak
% \vspace{-80pt} 
\end{samepage}
% \newpage

 

% \end{IEEEbiography}



% \vfill

\end{document}